# Economics of In-Space Industry and Competitiveness of Lunar-Derived Rocket Propellant

Author:

Philip T. Metzger (corresponding author)

Florida Space Institute, University of Central Florida, 12354 Research Parkway, Suite 200, Orlando, FL 32826; philip.metzger@ucf.edu.

## Abstract

Economic parameters are identified for an in-space industry where the capital is made on one planet, it is transported to and teleoperated on a second planet, and the product is transported off the second planet for consumption. This framework is used to model the long-run cost of lunar propellant production to help answer whether it is commercially competitive against propellant launched from Earth. The prior techno-economic analyses (TEAs) of lunar propellant production had disagreed over this. The "gear ratio on cost" for capital transport, $G$, and the production mass ratio of the capital, $\phi$, are identified as the most important factors determining competitiveness. The prior TEAs are examined for how they handled these two metrics. This identifies crucial mistakes in some of the TEAs: choosing transportation architectures with high $G$, and neglecting to make choices for the capital that could achieve adequate $\phi$. The tent sublimation technology has a value of $\phi$ that is an order of magnitude better than the threshold for competitiveness even in low Earth orbit (LEO). The strip mining technology is closer to the threshold, but technological improvements plus several years of operating experience will improve its competitiveness, according to the model. Objections from members of the aerospace community are discussed, especially the question whether the technology can attain adequate reliability in the lunar environment. The results suggest that lunar propellant production will be commercially viable and that it should lower the cost of doing everything else in space.

## Primary Symbols

| | |
|---|---|
| $C_{\text{fin}}$ | Cost of financing |
| $C_K$ | Cost of developing and fabricating capital |
| $C_{\text{ops}}$ | Cost of operations over the life of the capital |
| $c_R$ | Reliability cost factor |
| $C_{\text{tr}}$ | Cost of transporting capital to the lunar surface |
| $c_X$ | Cost per kg of lunar propellant delivered to location X |
| $E_R$ | Scale of effort required to increase reliability |
| $f$ | Specific finance cost |
| $g$ | Surface gravity of Earth |
| $G$ | Gear ratio "on cost" for delivering lunar capital |
| $G_{Y-X}$ | Gear ratio from location Y to location X |
| IMF | Inert Mass Fraction |
| $I_{sp}$ | Specific Impulse |
| $L_K$ | Cost per kg of launching capital from Earth to LEO |
| $L_p$ | Cost per kg of launching propellant from Earth to LEO |
| $M_K$ | Mass of capital |
| $M_{p,X}$ | Mass of lunar propellant delivered to location X over the life of the capital |
| $R$ | As-built reliability of the capital |
| $R_{max}$ | Maximum achievable reliability of the capital |
| $R_0$ | Baseline reliability of the capital |
| $T_K$ | Transportation rate (cost per kg) of capital to the lunar surface |
| $x$ | Launch-normalized equipment cost |
| $\Gamma_X$ | Propellant use ratio for delivery to location X |
| $\Delta v_{Y-X}$ | "Delta-v" (from the rocket equation) from location Y to location X |
| $\zeta$ | Capital development and fabrication cost rate ($/kg) |
| $\zeta_D$ | Capital development cost rate ($/kg) |
| $\zeta_F$ | Capital fabrication cost rate ($/kg) |
| $\kappa$ | Specific capital fabrication cost |
| $\lambda$ | Specific labor cost |
| $\xi$ | Launch-normalized finance cost |
| $\tau$ | Specific capital transportation cost |
| $\phi$ | Production mass ratio |
| $\chi$ | Launch-normalized capital cost |
| $\psi_X$ | Cost ratio of lunar vs. terrestrial propellant at location X |
| $\psi_0$ | Pre-delivery cost ratio of lunar (on LS) vs. terrestrial (in LEO) propellant |
| $\omega$ | Launch-normalized operations cost |

1. **Introduction**

Mining lunar water ice to produce rocket propellant may be an early business in cislunar industry. An example customer is a firm that owns and operates geostationary spacecraft and purchases boost services using lunar-derived propellant to place their satellites in operational orbit faster and at lower cost. Many see this "cislunar water economy" as vital for space development because it will have the scaling power of Earth's consumer markets, unshackling space development from the slower pace of government-led space exploration. Several techno-economic assessments (TEAs) of lunar-derived propellant have made either positive [1–4] or negative [5–7] predictions for its near-term economic viability. Others have expressed skepticism whether lunar ice mining can ever be economically viable because reusable launch vehicles will lower the cost of launching resources into space directly from Earth. Charania and DePascuale wrote,

> The price per kilogram for delivery of propellant from the lunar surface to a GEO [geostationary orbit] satellite tug customer does not provide an attractive alternative as compared to launch from Earth. The price per kg to send propellant from the lunar surface to GEO is extremely prohibitive…[T]he results of this study indicate a significant challenge in competing with propellants delivered from Earth [5].

Jones et al. agreed,

> … lunar ISRU [in-situ resource utilization, i.e., lunar-derived] propellant is 97% more expensive than Earth-based propellant… [I]t is unlikely ISRU is a cost-effective approach to providing propellant in cis-lunar space in the near term [6].

On the other hand, Bennett et al. reassessed the Jones et al. study by improving the economies of scale of the lunar surface hardware and concluded,

> We have proposed four alternative architectures that…reduce the cost of propellant at the cislunar aggregation point to well below the cost of commercial transport posited in Jones et al… We have demonstrated that there is a large space of architectural choices and that one region can be an order of magnitude better than another. This supports the potential for lunar ISRU propellant to make a significant contribution to future large-scale operations in cislunar space [2].

The divergent TEAs are difficult to synthesize. Shisko reviewed them and wrote,

> …the wide range of quantitative results is a consequence of the different assumptions, parameters, and analysis methods used in the studies. Because of this, it is difficult to compare results from different studies on a truly equal footing, i.e., apples-to-apples [8].

Part of the confusion is because these TEAs were not really on the same topic: they were assessing four different technologies (strip mining, borehole sublimation, tent sublimation, and excavation with beneficiation). Some of them concluded that lunar propellant manufacture is not economic, but that was illogical because they did not study the economics of lunar propellant manufacture per se but only the economics of one possible technology operating in that sector. The studies sometimes made basic scaling errors or used inadequate estimating methods that skewed the results, or they made choices for transportation systems that would guarantee a negative outcome, as I discuss below. Another basic problem is that the studies implied that the economics are static or that they change over a timescale that is too long for decisionmakers to care about, so they studied only point-designs at a single moment in time yet made longer-term pronouncements on that basis.

I think this confusion arose because we skipped an important step: studying the economics of the in-space propellant economy as a sector before doing TEAs on the possible technologies. We need a "spherical cow model" before adding the hoofs and the horns. This paper will take that approach by writing the equations for the primary, empirically supported economic factors that will determine the cost of space-mined propellant versus Earth-launched propellant and studying how those forces will drive in-space economics. This should produce greater confidence than a focus on the technological details to decide whether space mining is commercially feasible or not.

This study has the following goals. First, it will identify parameters that give insight into in-space industrial development. Rather than simply the costs of capital and labor, it recognizes the large role of mass and the rocket equation, so it focuses on mass-specific capital and mass-specific labor that have been normalized by launch cost, and it incorporates gear ratios from orbital dynamics. Second, this paper will assess the long-run trends in these metrics according to four economic factors: the cost of reliability, the experience or learning curve, economies of scale, and economies of scope. Third, it will try to provide a clear answer to the question whether lunar ice mining can be economic or not as a sector, and more narrowly, whether the decreasing cost of launch from Earth will inherently drive space resources out of business. Fourth, it will look at some of the details in the prior TEAs to try to explain why and how they disagreed. Fifth, it will hopefully create insight into what space companies need to do to make lunar mining more profitable. Sixth, it will show that policymakers must not ignore lunar resources because their economic potential will become disruptive to the status quo.

## 2. Assumptions and Ground Rules

For simplicity and clarity, some major issues will need to be ignored. This paper will focus only on liquid oxygen/liquid hydrogen (LOX/LH2) propellant, although SpaceX is building their transportation architecture around LOX/methane. Methane could be brought from Earth even while LOX is sourced from the Moon. LOX is 89%wt of the stoichiometric mass of water and as low as 80% of the mixing ratio in rocket engines, and an 11% to 20% error is not bad for a spherical cow model so this analysis is still valid. It is also possible to manufacture methane and other space-storable propellants like hydrogen peroxide using space resources, and some companies are already working on the technologies. Furthermore, if LOX/LH2 turns out to be economically superior when using lunar resources, it is not too late for companies to switch to these propellants for many future applications. On the other hand, the United Launch Alliance and Blue Origin have already focused on future use of lunar-sourced LOX/LH2. Introducing architectural variations would make the model too complex for the present purposes.

This paper will also ignore the possibility of transportation architectures using propellant depots in cislunar space [9–12]. The business case for depots is broader than just propellant sale and would complicate the analysis. For specificity this paper will focus on the concept of launching one load of propellant off the Moon at a time to provide a boost to a spacecraft then returning to the Moon for another load [13]. The model predicts cost savings from economies of scale in future years as the propellant market grows, and those economies would likely be achieved because companies shift toward building propellant depots with their broader scope, greater capital infrastructure, and better scaling, but exactly how these economies of scale and scope are achieved does not need to be explicit in this type of model and would introduce too many variables that make the analysis less certain.

This paper will furthermore ignore the many architectural choices for cislunar space transportation in addition to propellant depots, including the use of aerobraking, non-chemical propulsion technologies, mass drivers or sling launchers from the lunar surface, etc. Solar electric propulsion (SEP) will be included but only as an example of a high specific impulse system to show how this affects the economics. The details of SEP or other propulsion systems would introduce too many complicating variables. The goal of this paper is not to analyze architectures but only to study the economic feasibility of lunar resources using the simplest cases.

This study will also restrict its focus to the costs to produce and deliver propellant, not the pricing of the propellant and resulting profit, which involves additional economic dynamics.

Additionally, this paper will ignore the possible role of humans on the Moon working in this industry. The Jet Propulsion Lab's recent Robotic Lunar Surface Operations 2 study [14] developed lunar ice mining architectures in which humans visit periodically for short stints only to repair hardware. That may be a good model if the cost savings from repairing rather than replacing hardware is greater than the extra expense. There are companies developing business cases that include humans on the Moon, whether for tourism or industry, and national space agencies may pay the expense of humans on the Moon for national goals and exploration. It would be strategic for national space agencies to focus on developing lunar surface industry as one of the activities for their lunar astronauts. Although I am an advocate for humans working in lunar surface industry, including humans in this analysis would complicate it. To justify the cost of habitats, life support systems, construction systems for radiation shielding, higher safety standards in transportation, etc., there must be a larger business case than just propellant mining, and that introduces many additional variables. Instead, this paper will assume pure teleoperation of the robotics with modularized assets that can be replaced by teleoperation recognizing that as the industry grows there will be increasing opportunities for synergy between human and robotic lunar activities.

Autonomy of the robotics will therefore be important for the development of the industry. Some initial level of autonomy is needed for smooth operation despite the 3-second round trip time delay for communications with Earth, then loosely supervised or full autonomy can be phased-in over time for greater labor efficiency. This is discussed in Section 9.2.

This model assumes that each of the proposed lunar propellant production systems will perform as described by their individual TEAs. The technologies are mostly at low technological maturity (some that are at mid-range technological maturity are mentioned in Section 6.6) and there is risk that some may not work or may not perform as well as expected. On the other hand, ongoing innovation could improve performance to better than currently expected. There is additional risk derived from uncertainties in the geology and physical state of the lunar ice. Missions to the Moon to study the resource are already planned, and many groups globally are working to mature and test the technologies, so the technology assumptions in this model can be replaced by clarity over time.

## 3. Initial Considerations

### 3.1. *Plausibility in Physics*

Because of constraints in physics, the best payload mass fraction for conventional rocket technology launching off the Earth to GTO is about 2%. For launching off the Moon with its weaker gravity to GTO, the payload mass fraction can be about 48%, or 24 times higher. If this were the only difference between the Earth and the Moon, lunar propellant would be 24 times cheaper than Earth-launched propellant in GTO. However, on Earth it is trivial to acquire the raw materials that go into rocket propellant (e.g., water), whereas on the Moon they must be mined as ice in an extremely harsh environment. The Moon's capital (including its launch vehicle to carry propellant off the Moon) can be made on Earth and transported to the Moon, so the Moon does not need its own industrial supply chain, but there will be capital transportation costs. Humans can teleoperate the rockets and robots from Earth, so the Moon does not need its own workforce. (See the Discussion section about teleoperation.) However, operations will be less efficient when it comes to troubleshooting and repairing broken hardware with no humans on-site. With fewer successful repairs there will be more replacement of hardware, which adds to the capital acquisition and transportation costs. Maybe the lunar capital can be designed for better modularity with telerobotic replacement to minimize the extra mass, but still it must be designed to operate with adequate reliability in a harsher environment than Earth.

The question whether lunar propellant can be competitive comes down to this: can we (1) transport capital to the Moon, (2) tele-support its operation on the Moon, and (3) work with difficult raw materials in a harsh environment, with a total economic penalty that is less than a factor of 24 so it does not eat up the entire positive margin afforded by the physics? A factor of 24 is quite high (that is, a 2,300% cost increase over building and launching rockets from Earth), so there is no *a priori* reason to reject its plausibility.

*3.2. Comparative Advantage and Environmental Externalities*

Before beginning it is important to expose a common fallacy: the belief that lunar-derived propellant needs to be cheaper than Earth-launched propellant to be competitive. It needs only a comparative advantage, not an absolute advantage, since launching rockets entails a significant opportunity cost and buying propellant in cislunar space from a lunar mining operation may be comparatively less expensive than losing the opportunity to launch a more lucrative payload. In general, Earth-launched propellant does not compete against lunar propellant; it competes against the value of other payloads that can be launched from Earth.

This may not be relevant in the near or mid-term for a company like SpaceX, which is developing a super-sized rocket and expects a glut of launch capacity so they may want to launch the propellant themselves. However, it will still be important for companies competing with SpaceX and it becomes strategically crucial for SpaceX in the longer term. This paper is not based on SpaceX's plans, but they are part of the context for assessing lunar resources so the reader may benefit from a brief overview.

SpaceX's overarching goal is to build a city on Mars. Founder Elon Musk set the timeline at one million settlers on Mars by 2050 [15]. He said, "I'm only personally accumulating assets in order to fund this" [16]. Once people and manufacturing equipment start landing on Mars, it will take years of buildup before a city can break even financially, so the effort must be financed at extreme cost. Musk has stated this Martian city may cost between $100B and $10T [17], possibly beyond his ability to finance so every advantage is needed. He said, "ultimately this will be a huge public-private partnership" [16]. Supporting a population of a quarter million to a million people who cannot yet fully support themselves will be expensive, so accelerating the shipments of capital to make Mars self-sustaining sooner is highly leveraged to reduce net costs.

The time-value of cargo to Mars also includes intangibles such as extra security for the population by having more industrial capacity on-site sooner, creating progress faster before political opposition can galvanize on Earth (a benefit from the perspective of the investors), and seeing the project completed in the lifetimes of the investors.

As a hypothetical illustration, assume that lunar propellant has a cost in LEO that is 8% more expensive than launching the propellant from Earth. The modeling discussed below shows that, if SpaceX applies 50% mark-up for profit in their launch price for the lunar capital, that increases the cost of the lunar propellant by only 16% (because launch cost is only one of the cost drivers). If the lunar mining company marks up their sale price another 12% for their own profit, then SpaceX can purchase the propellant at 40% higher cost ($1.08 \times 1.16 \times 1.12 = 1.40$) than launching it themselves. Thus, by launching one load of capital in lieu of one load of propellant, SpaceX obtains a net 50% - 40% = 10% profit from the deal even though they could have launched the propellant at lower cost. The lunar capital then produces on the order of 100x more mass of propellant than the mass of capital (discussed below), so SpaceX can buy the additional propellant and use their freed-up launch capacity to send more capital to Mars. If the extra loads of capital each create only 20% annual return in value on Mars, which compounds to 48% per synodic period (Mars departure windows from Earth), then sending a load of capital one synodic period sooner is an immediate net gain even if the lunar propellant were 40% more expensive (48% - 40% = +8%). This benefit compounds continuously because the extra capital on Mars produces value through many years, and simple modeling shows this can dramatically cut the cost of Mars settlement. Capital on the Moon enables production of rocket propellant; capital on Mars enables production of goods that Mars residents consume plus materials to make more capital, so SpaceX avoids shipping costs. In both cases, the capital results in a greater mass of goods than the capital itself, which enables trade so comparative advantage is a factor.

Also, lunar propellant mass is obtained higher in Earth's gravity well, so staging for Mars can be done in Moon Distant Retrograde Orbit (DRO) instead of LEO (discussed below), which can reduce the propellant cost by 30% compared to the cost of bringing all the propellant from the Moon to LEO for refueling.

Furthermore, the launch rate in Musk's Mars settlement plan is perhaps 7x higher [18] or more [19] than rates predicted to cause significant damage to Earth's ozone. There are also concerns with carbon black emissions of hydrocarbon-fueled rockets affecting the thermal budget of the upper atmosphere (this effect with methane has not been quantified, yet) [20,21]. These concerns are a risk to Mars settlement because governments might limit the launch rate or might internalize the external costs by imposing environmental impact fees that drive up the launch cost, erasing any cost advantage of terrestrial propellant. The calls for this have grown exponentially in the past three years [22–27]. Reducing the launch rate by up to 9x by using lunar propellant will therefore reduce the political and environmental risks of a Mars settlement campaign. For these reasons, even if SpaceX expects a glut of launch capacity, it is strategic to plan on using lunar resources. Understandably, proponents of Mars settlement will reject putting the complexity of lunar mining into the critical path to Mars, but if other companies (and/or governments) develop lunar propellant at their own risk so it becomes available for Mars settlement, then failing to take advantage of lunar propellant would only make Mars settlement more costly with higher risk. Furthermore, this paper will show that lunar propellant can gain an absolute advantage even without these considerations.

Of course, lunar propellant does not rely on Mars settlement as its only business case. There are other space transportation companies like the United Launch Alliance (ULA) and Blue Origin that have identified lunar propellant as a near-term business strategy in cislunar transportation [28–30]. OrbitFab is already providing in-space water to NASA as a customer at the International Space Station and is signing on other customers for their services, including Astroscale's LEXI spacecraft that will be equipped with OrbitFab's refueling port. Daniel Faber (CEO of OrbitFab) points out that in the past 10 years about 200 spacecraft worth an aggregate $100B had to be discarded in space due to running out of propellant. There are also geopolitical considerations that should create government customers for in-space propellant at higher than launch market prices [29]. Listner [31] recently argued the geopolitical importance and urgency in developing lunar resources. The business risk for a lunar propellant firm would be lowest if they gain an absolute advantage over the Earth-launched propellant, but these broader considerations suggest they could weather a short-term glut of launch services even without it.

## 4. Model Development

### 4.1. Basic Approach

A complication arises because the amount of product sold, both terrestrial and lunar, affects the economy of scale and learning curve for each business, but the amount sold by each depends on relative pricing that differs with location across cislunar space and depends on other economic factors. Parsing the market demand by location would be too complicated for this simple model, so instead the demand for both rocket launches and in-space propellant are applied exogenously using three cases: one with optimistic growth of the launch market, the second with moderate growth, and the third pessimistic. The analysis assumes that the launch market and the in-space propellant market are related: when launch rates increase, then in-space activity and need for propellant increases in rough proportion. The launch costs and the lunar industry costs then develop independently based upon their own market models through learning curves and economies of scale and scope. I attempted to choose reasonable index parameters for each of these effects based on a thorough literature review given in the Supplementary Material. The model then calculates the costs of lunar and terrestrial propellant for "year 1" (the first year of product sales) and predicts which one's cost will drop faster per those index parameters over the next thirty years. This will indicate whether the competitiveness of lunar propellant may change over a timescale that is important to policy planners. For example, if lunar propellant is predicted to be more costly than terrestrial propellant in the first year of commercial sales, but the modeling shows it should become much less costly than terrestrial within only five or ten additional years, then it would be in a national space agency's interest to continue developing the technologies to spin them off to commercial firms to provide this service when the agency needs them. The agency may gain deep cost savings for (e.g.) Mars missions ten or twenty years in the future despite the poor cost comparison in the first year of commercial sales.

### 4.2. Cost Equation

This model focuses on Long Run Average Cost (LRAC) so appropriate averaging methods are employed. The average cost ($/kg) for lunar-derived propellant when the product is sold at the lunar surface may be approximated as,

$$c_{LS} = \{C_K + C_{tr} + C_{ops} + C_{fin}\}/M_{p,LS}$$
$$= \kappa + \tau + \lambda + f \quad (1)$$

where $C_K$ = cost of capital development and fabrication, $C_{tr}$ = cost of transporting the capital from Earth to the lunar operations site, $C_{ops}$ = operating cost over the life of the hardware, $C_{fin}$ = finance costs, and $M_{p,LS}$ = the total mass of propellant (p) produced at the lunar surface (LS) over the operational life of the hardware. Finance costs could have been included where incurred as part of the capital acquisition, transportation, and labor, but it was written separately to provide some insight. For a robotic lunar industry where humans operate the hardware telerobotically from Earth, $C_{ops}$ is mainly just labor with overhead. The purchase of telecommunications bandwidth between the Earth and the Moon is not a separate cost center because it is counted as part of the overhead on labor. The labor supply is assumed to have a constant wage rate and to scale inelastically as required to match the propellant demand. The aerospace workforce moves fluidly between companies so both the lunar propellant industry and the competing Earth-launched propellant business share the labor pool and wage rate. Spares procurement has been included in capital costs rather than operations costs since optimizing reliability entangles the amount of spares with the cost of capital (see Section 4.3). Energy is not a separate cost center because the individual TEAs for lunar ice mining always include energy generation equipment as part of the capital and the labor to operate that equipment is part of the labor. The (product mass-) specific capital, labor, transportation, and finance costs are $\kappa$, $\lambda$, $\tau$, and $f$.

Transporting the product from LS to other locations in cislunar space is considered here part of the firm's business, so transportation hardware such as a reusable lunar lander (RLL) and possibly a space tug or orbital transfer vehicle (OTV), their fabrication, initial delivery to point of operation, operational costs, and associated finance costs are already included in $\kappa$, $\tau$, $\lambda$, and $f$, but the act of transporting the product in those vehicles uses up a portion of the product – often the vast majority of it – so this requires an additional factor that does not occur in terrestrial economics. Simplistically, the mass of delivered product becomes,

$$M_{p,X} = M_{p,LS}\left[(1 + \text{IMF})\exp\left(-\frac{\Delta v_{LS-X}}{g\,I_{sp}}\right) - \text{IMF}\right] = M_{p,LS}/G_{LS-X} \quad (2)$$

per the Tsiolkovsky rocket equation, where $M_{p,X}$ = the mass of propellant at location X after transport, IMF = the inert mass fraction of the spacecraft, $\Delta v_{Y-X}$ is the delta-velocity from location Y to location X according to orbital dynamics, $G_{Y-X}$ is the "gear ratio" from Y to X, $I_{sp}$ = the specific impulse of the rocket propellant, and $g$ =9.8 m/s is Earth's surface gravity, which serves to define the customary units of $I_{sp}$. The notation X and Y can be replaced by LS = lunar surface, LEO = low earth orbit, GTO = geostationary transfer orbit, GEO = geostationary orbit, EML1 = the Earth-Moon Lagrange point 1, LLO = low lunar orbit, DRO = Moon distant retrograde orbit, etc. For an OTV and an RLL hauling rocket propellant or capital sequentially, the gear ratios for each are multiplied together. To account for the propellant the RLL and OTV need to return from their points X where they unload the goods back to their points of origin Y, replace IMF with the effective IMF,

$$\text{IMF}' = \text{IMF}\, \text{Exp}\left(\frac{\Delta v_{X-Y}}{g\, I_{sp}}\right) \quad (3)$$

The gear ratio in orbital dynamics is the ratio of the mass of hardware and propellant *before* versus *after* moving from location Y to location X per the rocket equation. If you want to deliver 1 kg to location X, you need to start with $G_{Y-X} \times 1$ kg at location Y. The cost of propellant at location X becomes,

$$c_X = (\kappa + \tau + \lambda + f)\, G_{LS-X} \quad (4)$$

$G_{Y-X}$ is a mass ratio through Eq. 2, but it can be interpreted as a cost ratio through the price per mass of propellant. Generalizing it as such enables analysis of architectures using different and mixed transport systems with varying base costs, propellants, reusability, etc., as discussed below.

The condition for lunar derived propellant to have lower cost than terrestrial propellant at location X is,

$$(\kappa + \tau + \lambda + f)G_{LS-X} < L_p G_{LEO-X} \quad (5)$$

where $L_p$ is the (mass-)specific launch cost of propellant from Earth to LEO in \$/kg. The cost of manufacturing and handling propellant on Earth and the operations cost of propellant transfer in-space are negligible, so the right-hand side of Eq. 4 is only the transportation cost.

Capital equipment costs for spaceflight hardware are usually estimated by mass, so $\zeta_D$ = development cost of equipment per kg (mass of the equipment, not of the product it produces), and $\zeta_F$ = fabrication cost of equipment per kg. For the first hardware set, the cost is $C_K = (\zeta_D + \zeta_F)M_K$, where $M_K$ = the mass of the equipment, but for additional production including spares and replacements it is $C_K = \zeta_F M_K$. For now, we write $\zeta = \zeta_D + \zeta_F$, but we address the ratio of spares to originals below. Then, $\kappa = \zeta M_K/M_{p,LS}$.

The capital's transportation rate (cost per kg) is $T_K = L_K\, G_{K,LEO-LS}$ and the (product mass-)specific transportation cost is $\tau = L_K\, G_{K,LEO-LS}\, M_K/M_{p,LS}$. The launch cost $L_K$ and the gear ratio $G_{K,LEO-LS}$ refer to the transportation system that transports capital. This could be different than the transportation system that launches the competing terrestrial propellant and delivers it to location X. We define the capital's "gear ratio on cost,"

$$G = \frac{L_K \, G_{K,\text{LEO-LS}}}{L_p} \tag{6}$$

which reduces to the ordinary "gear ratio on mass" (as we could call it) when a common transportation system is used to launch and deliver both capital and propellant, $L_K = L_p$.

We normalize the equation by $L_p$ since it is a convenient cost standard. Substituting the capital and transportation terms into Eq. 5,

$$\left\{ \frac{\zeta M_K}{L_p \, M_{p,\text{LS}}} + \frac{G \, M_K}{M_{p,\text{LS}}} + \frac{\lambda}{L_p} + \frac{f}{L_p} \right\} G_{\text{LS-X}} < G_{\text{LEO-X}} \tag{7}$$

which becomes

$$[(x + G) \, \phi^{-1} + \omega + \xi] \cdot \Gamma_X < 1 \tag{8}$$

where we introduced dimensionless parameters: *production mass ratio*, $\phi = M_{p,\text{LS}}/M_K$; *launch-normalized equipment cost*, $x = \zeta/L_p$; *launch-normalized operations cost*, $\omega = \lambda/L_p$; *launch-normalized finance cost*, $\xi = f/L_p$; and *propellant use ratio*, $\Gamma_X = G_{\text{LS-X}}/G_{\text{LEO-X}}$. For brevity we may refer to these quantities without writing "launch-normalized" every time, the context making it clear. If $x$, $\omega$, and $\xi$ are set to zero, the remaining term $G\phi^{-1}\Gamma_X$ is the inverse of the "propellant payback ratio" defined by Pelech, et al [7]. The technology parameters are $x$, $\phi$, and $\omega$. The transportation parameters are G and $\Gamma_X$.

We further define the (*launch-normalized*) *capital cost*,

$$\chi = (x + G) \, \phi^{-1} \tag{9}$$

which includes both fabrication and transportation of the capital and is specific to the mass of product rather than the mass of capital. The *cost ratio* of lunar vs. terrestrial propellant at location X is

$$\psi_X = (\chi + \omega + \xi) \, \Gamma_X \tag{10}$$

or we can write the *pre-delivery cost ratio*, $\psi_0 = (\chi + \omega + \xi)$. Lunar propellant has absolute advantage at X when $\psi_X < 1$. Sometimes it is more convenient to plot $\psi_0$ so the absolute advantage is when $\psi_0 < 1/\Gamma_X$. The rest of this paper will study how these dimensionless parameters behave.

*4.3. Cost of Reliability*

Low launch costs help make lunar propellant more competitive by making it inexpensive to replace failed hardware, so ultra-high reliability is not needed for lunar mining where high reliability would be expensive to achieve. Empirical data show that the hardware cost rate $\zeta$ scales exponentially with reliability. A common saying is that NASA pays 90% of the cost for the last 5% of reliability (the numbers vary with the telling). For space agencies, the politics of mission failure drive even higher reliability than economics do. To help alleviate this, NASA has been implementing commercial provider programs that create some political insulation, including Commercial Lunar Payload Services (CLPS), which aims to put higher risk but lower cost uncrewed landers on the Moon. This is expected to increase the science return on the dollar despite a higher failure rate [32] due to the non-linearity of the reliability cost function. On the other hand, a commercial operation needs to attract customers, who will want assurance of reliable boost or reliable delivery of propellant. Here the assumption is that a commercial space mining company will allow lower reliability at the hardware element or subassembly level as a cost savings but will manage spares and repairs to provide high reliability in delivery of the final product.

Mettas [33] wrote a cost-reliability model by fitting to empirical cost-reliability data,

$$c_R = \exp\left[(1 - E_R)\frac{R - R_0}{R_{\max} - R}\right]$$

(11)

where $c_R$ is the reliability cost factor that is multiplied onto $\zeta$, $R_0$ is the baseline reliability corresponding to the baseline cost, $R$ is the reliability as built, $R_{\max} \cong 100\%$ is the maximum achievable reliability. $R_0$ is the baseline reliability corresponding to the baseline cost model. The baseline is built using baseline quality components and baseline design resilience (internal redundancies and robustness of the internal architecture) and is subjected to a baseline amount of testing, so as-built reliability and cost can go either higher or lower than baseline. $E_R$ is a parameter between 0 and 1 that scales the effort of improving reliability. Stancliff, et al. [34,35] used $E_R = 0.5$ and $E_R = 0.95$ for lunar rovers and obtained similar results in each case. Here, the model uses $E_R = 0.5$.

Jones et al. [6] assumed 10% of the mass of lunar propellant production hardware must be replaced each year over a 10-year operational life, which is $R = 0.50$ (half of the hardware that is produced including spares fails during the baseline period). This can be checked using the Feasibility of Objective Technique from Military Handbook (MIL-HDBK)-338B [36] benchmarked by spacecraft data having $R = 0.92$ for 10-year operation provided by Figure 3 of Yang et al. [37] This estimates that a lunar strip mining technology using nuclear power as described by Bennett et al. [2] would have mass-weighted $R = 0.36$ for 10-year operation with an annual spares mass of 28%, much higher than the value of Jones et al. This assumes a failed element must be replaced entirely, so if we assume most failures occur in modules that are 10%wt apiece while others like RLL engines must be replaced in whole, this improves the mass-weighted reliability to $R = 0.72$ and the annual spares mass to 3.9%. This reliability value is dominated by the technological immaturity of the nuclear reactor and chemical processing (water cleanup and electrolysis) elements and the excavators operating full-time in harsh lunar dust. Replacing the nuclear reactor with the solar power system described by Sowers [4] improves the mass-weighted reliability to $R = 0.76$ with annual modularized mass of spares at 3.2%. For the tent sublimation technology described by Sowers [4], this technique estimates base reliability $R = 0.80$ with annual modularized mass of spares at 2.6%. The model will use $R = 0.78$ for both methods. Achieving this degree of modularity with robotic replacement is crucial for these annual spares masses lower than the Jones et al. estimate. Additional cases will be run with $R_0 = 0.20$ to $R_0 = 0.92$ to test sensitivity.

The total mass of capital that must be fabricated and transported to the Moon includes spare modules and spare system elements. This scales as $1/R$ in the long-run average, so $\zeta = \zeta_D + \zeta_F/R$. The cost of developing, fabricating, and transporting hardware becomes,

$$C_K + C_{tr} = \left(\zeta_D + \frac{\zeta_D}{R}\right) M_K \exp\left[(1 - E_R) \frac{R - R_0}{R_{max} - R}\right] + \frac{M_K}{R} T_K \tag{12}$$

where $T_K = L_K G$. This can be minimized by varying $R$. Fig. 1 plots example cases for four transportation rates. The minima are marked by solid dots, which define the cost-optimized reliability $R_{opt}$. $R_{opt}$ occurs at lower $R$ when $T_K$ is lower, so the cost of equipment fabrication drops when the cost of launch drops.

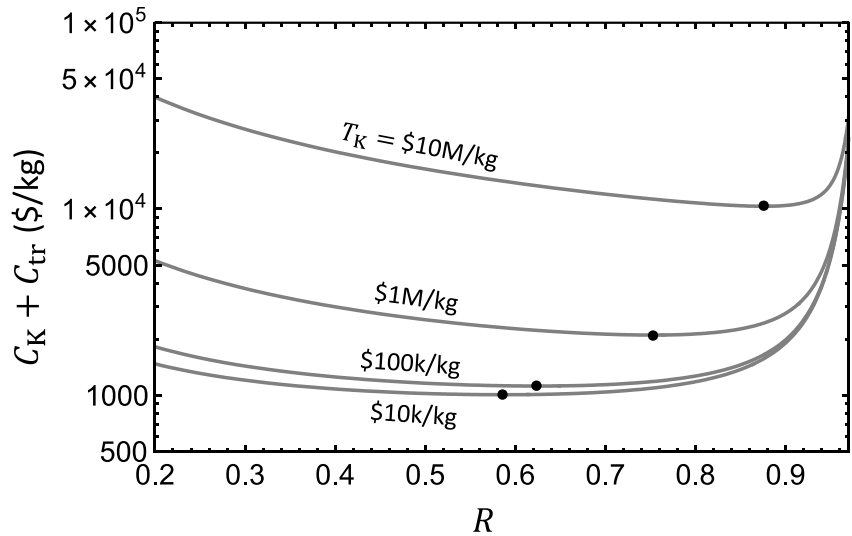

**Figure 1.** Sum of capital acquisition and transportation costs vs. reliability for four transportation rates to the lunar surface. Solid dots are the optimum reliability for each curve. Calculated from baseline reliability $R_0 = 0.78$.

Fig. 2 shows $R_{opt}$ vs. $T_K$, and fig. 3 shows optimized reliability-cost factor $c_R$ versus $T_K$. The parameter choices appear reasonable: if a lunar mining rover with this baseline reliability were brought up to 96% reliability through improved design and extra testing, it would increase the development and fabrication cost by a factor of 8. However, lowering the reliability to optimize for low-cost transportation results in only 12% savings. The main finding is that it is unnecessary to make costly improvements to the hardware to gain higher reliabilities typical of NASA exploration missions, so the environmental harshness of mining in lunar dust will not drive costs exceedingly high.

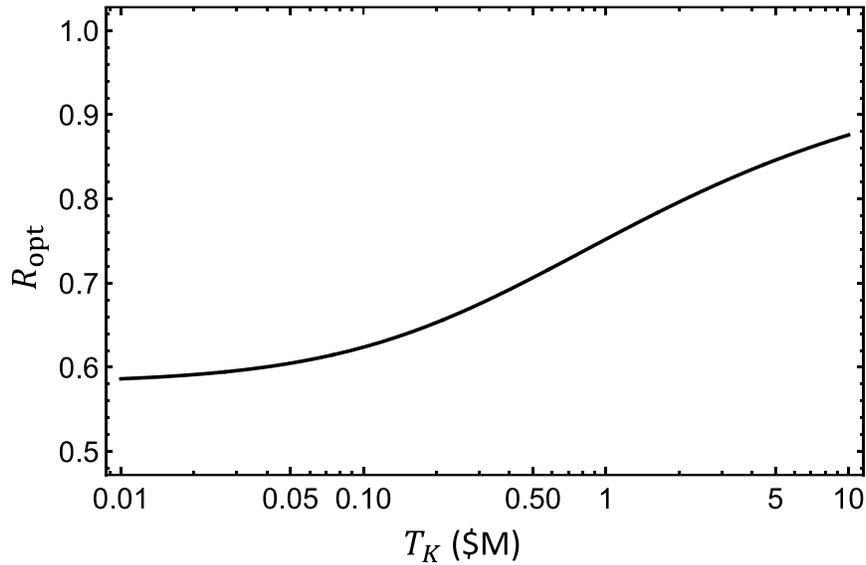

**Figure 2.** Cost-optimized reliability $c_R$ vs. transportation cost $T_K$ for $R_0 = 0.78$.

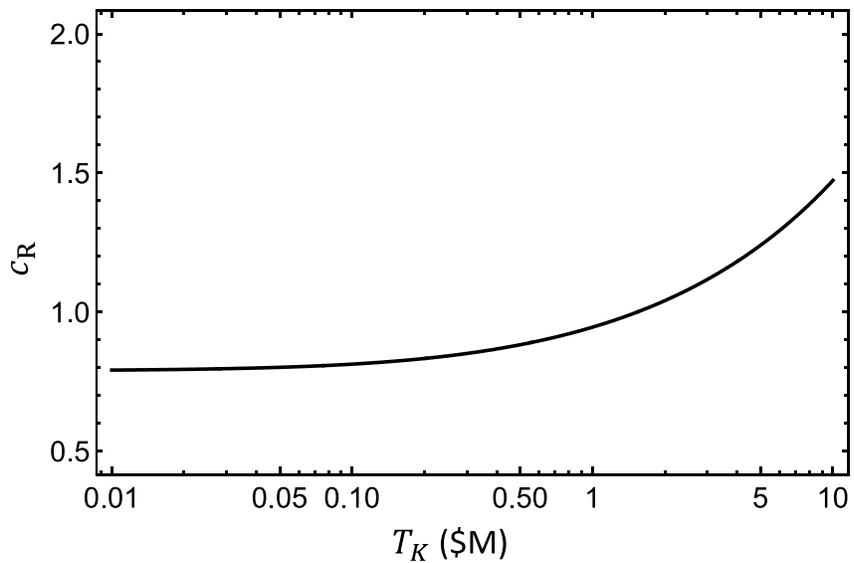

**Figure 3.** Optimized Reliability Cost Factor vs. Transportation Cost for $R_0 = 0.78$.

*4.4. Experience Curve*

There is a deep economics literature on industrial scaling relationships that can be applied to the unique conditions of in-space industry, including the experience curve (or learning curve), economies of scale, and economies of scope. The Supplementary Material provides a more detailed rationale of the model choices resulting from the literature [38–67].

Wright's Law for the experience curve is,

$$c_{WL} = \left(\frac{S(t)}{S(0)}\right)^{\text{Log}_2(b)}$$

$$S(t) = \int_0^t P(t')dt' + S(0)$$

(13)

where $c_{WL}$ is the factor that is multiplied onto the baseline production cost, $P(t)$ is the production rate, $S(t)$ is the total production up to time $t$, $S(0)$ is the total production up to $t = 0$, which is the time of the baseline production cost, and $b$ (a positive number between 0 and 1) is the progress ratio. A common value in industry is $b = 0.80$ with broad variation. Here, $b = 0.75$ seems conservative since this is a new industry with more room to learn in the early decades than the average terrestrial industry, and great advances are expected in robotic autonomy for reducing teleoperator labor. The value of $b$ will be varied to test parameter sensitivity.

*4.5. Economies of Scale*

Economies of scale (EOS) are distinct from the learning curve [41,54,58]. Haldi and Whitcombe [48] derived the relationship on industry data,

$$c_{EOS} = \left(\frac{X}{X_0}\right)^{a-1}$$

(14)

where $c_{EOS}$ = the factor on the unit production cost, $X$ = production capacity, $X_0$ = baseline production capacity corresponding to baseline unit cost, and $a$ is an empirical parameter often close to 0.6. There is a good argument (see Supplementary Material) that lunar ice mining will not reach the optimally maximum scale economy within 30 years even assuming the optimistic market demand model. Nevertheless, for conservatism we can implement a limit to EOS in the firm-level costs where a maximum is more likely to occur, as opposed to the cost of parts in the supply chain where market growth will not likely exhaust EOS over the modeled period. Data from terrestrial mining, biofuels production, farming, and other industries contributed to the model choices here (see Supplementary Material). The normalized equipment fabrication and labor costs with EOS become,

$$x' = x \left(\frac{X_{\text{props}}}{X_0}\right)^{(a-1)/2} \left(\frac{\text{Min}[X_{\text{props}}, X_{\text{max}}]}{X_0}\right)^{(a-1)/2}$$

$$\omega' = \omega \left(\frac{\text{Min}[X_{\text{props}}, X_{\text{max}}]}{X_0}\right)^{(a-1)}$$

(15)

where $X_{\text{props}}$ = the production rate (kg/y) for propellants, $X_0$ = the initial propellant market correlating to baseline costs, $X_{\text{max}}$ = the optimally maximum production rate beyond which firm-level EOS ends, and the Min function selects the minimum of the two arguments. The EOS for equipment fabrication has been split equally into two terms: one for supply chain EOS and the other for firm-level EOS, the latter one limited by $X_{\text{max}}$. An example of a firm-level limit is when the payload faring diameter of the rocket that launches chemical processing reactors to the Moon is not large enough to accommodate larger reactors, and the firm has not yet innovated ways to assemble larger vats after lunar delivery. We can assume the operations cost shares the EOS limit with firm-level capital limits if the number of personnel scales with the mass of equipment. We will investigate cases with the conservative values $a = 0.6$ and $0.8$, and with firm-level EOS limits $X_{\text{max}} = 10$ and 20 t/day based on analogy to the biofuels industry. Supply-chain-level EOS will be modified further for economies of scope.

*4.6. Economies of Scope*

Economies of scope (SOE) is an effect from having overlap with other industries in the supply chain or other externalities, or from a firm developing multiple products that have commonality in capital assets and processes. Lunar propellant mining should benefit from supply chain SOE as other economic activities develop in cislunar space. A conservative estimate is that lunar metal production compared to lunar ice extraction will have $F_{\text{overlap}} = 50\%$ overlap in the supply chain technologies (rovers, telecommunications, landing pads, regolith excavating and handling, beneficiation, regolith heating technology, reusable lunar landers to deliver the product to the point of sale in cislunar space) so economies in the supply chain for ice mining will scale with 50% of the metal production rate.

Firm-level operations also share common processes (bringing strip-mined material to the processing plant and heating it for thermal extraction of ice, so it is already delivered and preheated for subsequent metal extraction), plus management, administration, legal support, and technical skills, so firm-level SOE should also exist.

In addition to metal production there are many examples of cislunar industries that may start after launch costs are adequately low and as robotic autonomy becomes more capable, but for conservatism only lunar metals will be included here. The results may underestimate the economic competitiveness and value of lunar propellant in the long run. Investors and policy planners should perform additional modeling that includes the broader SOE.

The normalized hardware and labor costs with both EOS and SOE become,

$$x' = x \left(\frac{X_{\text{props}} + F_{\text{overlap}} X_{\text{metals}}}{X_0}\right)^{(a-1)/2} \left(\frac{\text{Min}[X_{\text{props}}, X_{\text{max}}]}{X_0}\right)^{(a-1)/2} \left(1 - \frac{f_{\text{firmSOE}} X_{\text{metals}}}{X_{\text{props}}}\right)$$

$$\omega' = \omega \left(\frac{\text{Min}[X_p, X_{\text{max}}]}{X_0}\right)^{(a-1)} \left(1 - \frac{f_{\text{firmSOE}} X_{\text{metals}}}{X_{\text{props}}}\right)$$

(16)

The modeling will assume ad hoc that the lunar metals industry begins 10 years after the propellant industry with a production rate $X_{\text{metals}} = \beta X_p$ with $\beta = 0$ for $0 < t < 10$ years, linearly increasing to $\beta = 0.3$ during $10 < t < 15$ years, and $\beta = 0.3$ for $t > 15$ years. It will assume $f_{\text{firmSOE}} = 20\%$ matching the mid-value for the terrestrial examples cited above, and $F_{\text{overlap}} = 50\%$. These parameters will be varied to test sensitivity. The Supplementary Material provides more discussion of these choices.

### 4.7. *Launch Cost*

This spherical cow model does not assume a particular launch system. The following discussion is only to pick a useful range of values. Current launch cost to LEO on Falcon 9 rockets is about $L_0 = \$2,000$/kg. The claim is often made that the SpaceX Starship will achieve very low launch cost, perhaps as low as \$30/kg to LEO, with reduction caused in part by the new technology (the larger rocket) and in part by the faster launch cadence as the decreasing launch cost reaches tipping points that expand the scale of in-space activity. Both are examples of economies of scale and learning curve. Here we focus on the next 30 years and use the economies of scale and learning curve equations to fit this expected cost reduction. SpaceX had launched 171 Falcon 9 rockets by the end of 2021 with a payload capacity of ~22.8 t each, with 31 of the launches in 2021. For the simple modeling here, assume their annual up-mass $U_{\text{LEO}}$ increases exponentially

$$U_{\text{LEO}}(t) = U_0 \, \text{Exp}(t/\tau_L)$$

(17)

where $U_0 = 31 \cdot (22.8 \text{ tons})$ per year and $\tau_L$ will be determined. Applying EOS and Wright's Law to the initial cost,

$$L_p(t) = L_0[\text{Exp}(t/\tau_L)]^{a-1} \left[\frac{\int_0^t U_0 \text{Exp}(t'/\tau_L)\, dt' + S_0}{S_0}\right]^{\text{Log}_2 b}$$

(18)

where $S_0 = 171 \cdot (22.8 \text{ tons})$. With $b = 0.8$ for the typical learning rate and $a = 0.66$ for the industry-average EOS per Haldi and Whitcombe [48], this will achieve $L_{30} = \$30$/kg by $t = 30$ years if $\tau_L = 4.67$. But $\tau_L = 30/\text{Log}(U_{30}/U_0)$ so $U_{30} = 436,000$ t/y up-mass. This is equal to eight launches per day with the Starship 150 t payload capacity to LEO. This would be aggressive, but it is comparable to Musk's plan for settling Mars and is comparable to the rate that is probably needed to achieve $30/kg, so this was a fortuitous result. It suggests that the optimistically low launch cost is feasible, if the optimistically high launch rate also occurs. Musk stated there would 1000 Starships building up in LEO during each 26-month planetary alignment cycle, then departing to Mars together over a 30 day window, and that this process would continue over a period of 20 years [68]. To refuel 1000 Starships in LEO would require another ~8000 launches per 26 months for an average of ~11.4 launches per day average. The goal here is not to defend the scenario but to assess whether it could drive launch costs so low that lunar propellant cannot compete, since this is the extreme case and because it is affecting peoples' views on lunar resources. Alternative economic parameters that achieve the same outcome are $b = 0.7$ and $a = 0.85$. Either way, the parameters are reasonable to match Musk's scenario, and Eq. 18 may be approximated as an exponential decay of launch cost from $L_0$ to $L_{30}$.

*4.8. In-Space Propellant Market Model*

The goal here is not to predict the propellant market but to create an adequate model for testing whether lower launch costs can inherently out-compete lunar-derived propellant as the skeptics of lunar resources claim. A NASA-sponsored 2020 study by the Science and Technology Policy Institute wrote that lunar propellant will not be economically viable "if Starship meets its cost and performance targets" [69]. To disprove this, one must consider the range of future Starship prices as they continue dropping. A fair comparison will use a market size commensurate with the launch price.

Studies have predicted that the space commerce market should be price elastic [70,71] but the elasticity would not appear until launch prices have dropped below a threshold. Andrews and Andrews [72] wrote, "All available market data indicates that the space transportation market has both an inelastic and elastic portion" and that the elasticity threshold is $1320/kg (converted from lb to kg) in 2001 dollars, so $2200/kg in 2022 dollars. Ross et al. [18] estimated the elasticity threshold was $4400/kg (converted from lb to kg) in 2009 dollars, so $6100/kg in 2022 dollars, and used price elasticities of 1.5 (pessimistic), 2.0 (nominal), and 2.5 (optimistic) in their modeling. Kutter and Sowers [4,28] likewise argued that the market will be highly elastic when transportation costs are adequately low, and that in-space propellant storage and transfer will play a key role in these cost reductions.

Launch prices have decreased slowly since the 1990s, but new vehicles (Long March 3B, Dnepr, Proton-M, and Falcon 9), priced near or below the expected elasticity threshold, began flying in the late 1990s and early 2000s. As they gained market share, they accelerated the decline in global-average launch cost, which appears to have gone below the expected elasticity threshold around 2000 to 2010 [73–75]. The global annual number of orbital launches declined from the 1960s until about 2005 despite the slowly decreasing launch costs, indicating inelastic behavior as expected above the threshold, but it suddenly began rapid growth in 2005, which has continued unabated [76]. Before this change, from 1980 to 2005 the cumulative mass of objects in orbit increased at an average annual rate of ~140 t/y. This doubled to ~280 t/y starting in 2005, doubling again to ~550 t/y since 2019 [77]. A rough estimate is that the global-average launch cost decreased ~40% since 2005 when the annual launch mass doubled. This is consistent with price elasticity $E \sim 1.9$, close to the nominal value of Ross et al. [18], but if the trend since 2019 is real then $E \sim 3.8$. These figures are crude but are the best available.

The market demand is modeled as,

$$D(t) = D_1 \, \text{Exp}(t/\tau)$$

$$\tau = 30/\text{Log}(D_{30}/D_1)$$

(19)

Where $D_1$ and $D_{30}$ are the demand in years 1 and 30. $D_1$ is assumed equal to the initial production capacity of the lunar propellant business, based on the idea that the firm successfully markets its initial product. Values of $D_1$ for the various lunar propellant studies are given in Table A-1, Appendix A.

For $D_{30}$, this study will consider three scenarios: optimistic, moderate, and pessimistic markets. For the optimistic scenario we consider setting the demand in LEO to the amount of propellant that, if launched from Earth, would drive the launch cost to the low limit $30/kg per Section 4.7, which is $U_{30} = 436$ kt/y. This rate is consistent with Musk's scenario for Mars settlement. For cases using solar electric propulsion (SEP) from LLO to LEO (discussed below), the propellant produced at LS must be $D_{30} = G_{\text{LS-LEO}} \, U_{30} = 2.48$ kt/day. This is high but the modeling shows it is achievable if this size market actually exists.

For the moderate and pessimistic cases with much smaller market and modestly higher terrestrial launch cost, we consider setting $U_{30}$ to 10% and 1% of the optimistic value, respectively, or the equivalent of one Starship launch every 1.25 days and one every 12.5 days, respectively. For comparison, the Falcon 9 launch rate in 2022 was once every 11.8 days. In each case, $D_{30} = G_{\text{LS-LEO}} \, U_{30}$. If Starship launched at these slower rates, then per Eq. 18 the launch cost at 30 years would be $L_p(30) = \$119$/kg and $\$436$/kg to LEO, respectively.

These choices are checked for reasonableness by the market elasticities they imply. The optimistic market scenario is consistent with a price elasticity of

$$E = -\frac{\ln(U_{30}/U_0)}{\ln(L_p(30)/L_0)} = 1.58$$

(20)

Although this is the "optimistic" case, it is only a conservative elasticity, between the nominal and pessimistic values used by Ross et al. [18], and much lower than the elasticity currently indicated by the launch market since 2005. The optimism is only in the low launch cost, not in the resulting size of the market. This means that lunar propellant will gain smaller EOS and learning curve improvements in the model than might reasonably be expected. The moderate and pessimistic scenarios are consistent with $E = 1.52$ and $1.26$, respectively. These are all conservative tests for lunar propellant.

For our purposes it is not necessary to prove that the lowest launch cost and correspondingly largest market scenario will occur, only that if they do occur then lunar propellant can still compete, to answer the claims of skeptics. Nevertheless, the reader may wish to consider possible use-cases for such a large propellant market. The example previously mentioned is large-scale settlement of Mars, which is the primary goal of SpaceX. This would consume an annual propellant supply equal to the optimistic scenario. Musk's plan is to stage 1000 vehicles at a time in LEO, but an alternative to reduce crowding in LEO is to stage them in Moon Distant Retrograde Orbit (DRO) [78–80], Near Rectilinear Halo Orbit (NRHO) [81], Lunar Distance High Earth Orbit (LDHEO) [82–84] or even EML-1 [85]. Conte et al. [79] showed that DRO has dynamical advantages over LEO, first because it enables monthly Mars departures via Earth flyby for at least 7 of the months per 26-month synodic period all with reasonable $\Delta v$ and time of flight (ToF), and second because supplying the propellant from the Moon reduces the total propellant needed from LEO to Mars and it reduces the effective gear ratio of Mars transfer. Thus, DRO has net economic benefit over LEO.

Other activities contributing to an optimistic-scale propellant market may include supporting national exploration goals and defense activities, space tourism, developing multipurpose in-space infrastructure, providing in-space services, and developing other space resources. Propellant demand may include activities that are not yet commercially viable because national governments and some wealthy individuals are investing in the space economy in expectation that they will become viable. The prospects for these activities have been discussed elsewhere [28,69–71,87].

*4.9. Capital Delivery Gear Ratio*

The value of $G$ depends on the transportation architecture for delivering capital to LS. If an architecture like the fully reusable Lunar Starship architecture is used, commonly expected parameters predict that 15 flights may be needed to include refueling in LEO and LLO so that 150 t of payload can be landed on LS and all 15 Starship-like vehicles returned to land on Earth. Since the payload capacity to LEO for one such vehicle is also 150 t, the effective $G = 15$. Alternatively, if the Starship-like vehicle delivers capital only to EML-1 while an RLL (LOX/LH2, IMF=0.10) takes the capital from EML-1 to LS, then $G = 8.5$. If the RLL acts as the tug all the way from LEO, $G = 6$, which is the baseline case. This assumes both capital and terrestrial propellant are launched using the same rocket to LEO, so $L_K(t) = L_p(t)$ per Eq. 18. (Some of the TEAs used $L_K$ that is much more costly than $L_p$, giving terrestrial propellant an unfair advantage, as discussed below.)

*4.10.  Lunar Propellant Delivery*

In this model, delivery of the lunar propellant to the point of sale is performed by the lunar mining company using the RLL and OTV that are the firm's capital assets. Its fabrication and operations costs were included. Many architectures and technology alternatives are available that can lower the cost of delivery or increase the price of sale [3,88,89]. These include aerobraking around the Earth, multi-stage ascent and delivery vehicles, propellant tanks that are left in LLO to reduce the up/down mass to LS, and the use of stoichiometric fuel/oxidizer ratio, which lowers $I_{sp}$ but increases $\Delta v/\$$ if there are no other customers for the excess oxygen. Neglecting them here sets a higher bar for proving the competitiveness of lunar propellant, and lunar mining firms will find greater profit margin than predicted here by including these enhancements.

For the RLL, the model uses LOX/LH2 with $I_{sp} = 450$ s and IMF=0.10. We will consider some cases with high efficiency, low thrust SEP using molecular water propellant for transportation from LLO toward Earth. Such thrusters are being developed [90–93] and may have $I_{sp} = 1,000$ to 4,000 s. Here we use $I_{sp} = 2,000$. An alternative is to use an existing (Xenon or Mercury-based) electric thruster and launch the propellant from Earth to keep refilling it for a small cost, but this is not considered here. If SEP is used, the long transportation time to the point of sale is accounted as an increase in the cost of the propellant via a discount rate applied to the transportation time. LLO to LEO is assumed to take 1.5 years based on the SMART-1 transit time, and the times to intermediate locations are prorated linearly by $\Delta v$. Note that in this "spherical cow" model, SEP is used only as an example of high specific impulse alternatives to explore the general effects of varying the in-space transportation. Many other alternatives exist.

The baseline RLL as defined in Table A-1 has adequate payload capacity to deliver the entire annual product of propellant from LS to cislunar space assuming each roundtrip takes two weeks plus one week for maintenance between trips. It is also adequate to deliver the capital to LS during the buildup period.

### 4.11. *Cost of the Competing Terrestrial Propellant*

For the cost of terrestrial propellant competing against lunar propellant, we consider two architectures. The first is like the (expected) fully reusable Lunar Starship architecture; the second uses a Starship-type architecture only from Earth-launch to LEO then an OTV/RLL ($I_{SP} = 450$ s, IMF=0.10) to location X in cislunar space. The Starship-type architecture for *capital* delivery (X=LS) had $G = 15$, but in that case the payload was limited to 150 t whereas if the payload is the propellant itself and there are crossfeed lines to deliver propellant from the vehicle's tanks, extra refueling in LLO enables delivering 714 t to LS so the gear ratio improves to $G_{p,LEO-LS} = 10$. However, the gear ratio with an OTV from LEO to EML1 (reserving propellant for the OTV to return to LEO) and an RLL from EML1 to LS (reserving propellant to return to EML1), is better at 7. Although this is better for propellant efficiency, the modeling here found that the cost to develop the OTV/RLL (using the same cost model as the lunar capital but with high reliability since it will not operate on the harsh lunar surface) made it non-competitive against the baseline Starship in most cases. Therefore, the reusable Starship architecture is chosen as baseline in the model. The cost of terrestrial propellant scales down over time per the learning curve and economies of scale contained in $L_p(t)$ in Eq. 18, while gear ratios remain constant per the physics.

### 4.12. *Propellant Use Ratio*

The numerator and denominator of $\Gamma_X$ depend on the transportation architectures for delivering lunar and terrestrial propellant, respectively, as described above. After specifying the two architectures, $\Gamma_X$ can be mapped to locations in space. Two examples are given in Fig. 4. The lower the curve, the more competitive lunar propellant is at that location in space.

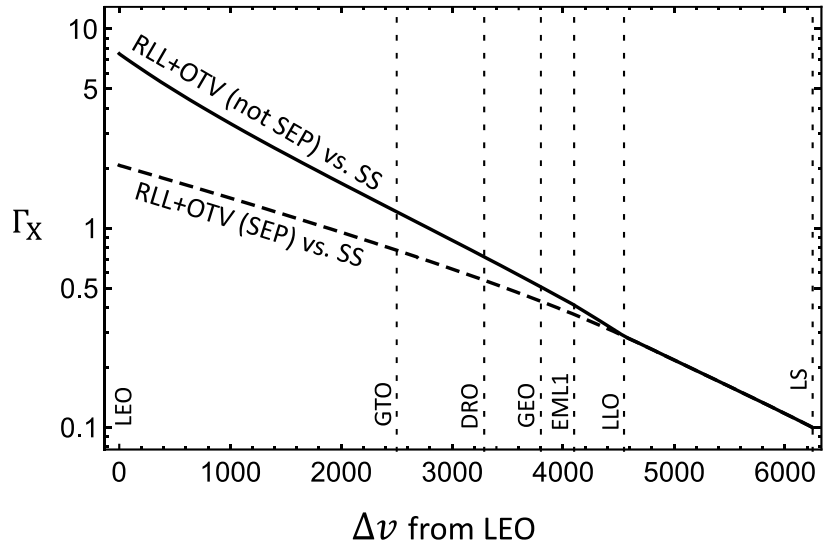

**Figure 4**. Delivery Cost Factor $\Gamma_X$ vs. $\Delta v$ from LEO for two combinations of transportation architectures. A lower curve means greater advantage for lunar propellant. OTV = Orbital Transfer Vehicle between LEO and EML-1, $I_{SP}$=450 s for non-SEP or $I_{SP}$=2000 s for SEP, IMF=0.10. RLL = Reusable Lunar Lander with hybrid propulsion, $I_{SP}$=2000 s (SEP) between EML1 and LLO or $I_{SP}$=450 s between LLO and LS, IMF=0.10. SS = fully reusable Starship-type vehicle $I_{SP}$=375 s, IMF=0.07. SEP = solar electric propulsion.

### 4.13. *Finance Cost*

For baseline, the discount rate starts at 21.7% following the Weighted Average Cost of Capital deemed minimally viable by Charania and DePascuale [5] considering risk and other factors. It decreases linearly to 12% at 30 years to represent the steady reduction of risk. The cost of developing and fabricating the equipment is assumed to be incurred uniformly through the years of the buildup period and the annual labor charge is also incurred each year during that period. That period is set to 5 years for the baseline. These expenses are financed since there is no revenue yet. Uniform payment/future value is used to calculate the accumulated debt by the end of the buildup period. The launch cost is incurred at the end of the buildup period, increasing the debt. Then present value/equal payment is used to calculate the necessary annual net income to retire the debt by the end of the operational life of the capital. Debt payment plus the annual operations cost is the necessary gross income, which determines the cost of the propellant. The total interest over both the buildup and operational periods divided by the total mass of product is the specific finance cost $f$. Dividing that by the launch cost gives the launch-normalized finance cost $\xi$.

## 5. Modeling Results

Lunar industry was modeled using the technology parameters in Table A-1 (Appendix A). The modeling is performed by (1) applying parameters for the specific case, (2) calculating $R_{opt}$ given those parameters and for the value of $L_K(t)$ in each year 1 to 30, and (3) calculating the total cost of lunar propellant at each location in cislunar space for each year using these parameters and $R_{opt}$. For comparison, the cost of Earth-launched water is calculated for each location in cislunar space for each year 1 to 30. Year 1 refers to the first year that product is manufactured and delivered, after the buildup period is complete.

The purpose of the baseline case is to investigate the economic relationships in the sector. No particular mining technology is specified. This case uses technology parameters that are round numbers and mid-range among the extant TEAs in Table A-1. The transportation parameters for capital, terrestrial propellant, and lunar propellant were not taken from those studies but were specified as described above to provide the best possible analysis of the overall sector, since some of the TEAs had made poor transportation choices. Uniform transportation parameters will also enable an apples-to-apples comparison between the TEAs. The baseline case will be considered in three market scenarios.

### 5.1. Baseline Case, Optimistic Market Scenario

The baseline/optimistic scenario results are shown in Fig. 5. The cost of propellant drops over time at every location in cislunar space. Lunar propellant drops faster than launch costs, so the region of space where lunar propellant is cheaper expands toward LEO. Lunar propellant has an absolute advantage down to DRO and GEO by year 1 and to GTO by year 5, almost reaching to LEO by year 30. This overturns the claim that decreasing launch cost makes lunar propellant less competitive.

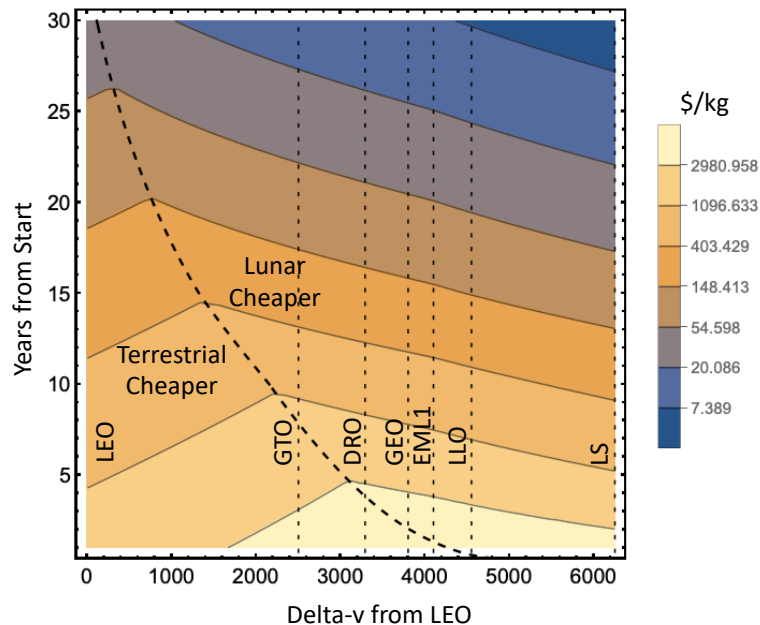

**Figure 5.** Cost of propellant in cislunar space over time for the baseline case in an optimistic market, either terrestrial-sourced or lunar-sourced, whichever is cheaper. Dark dashed line separates region where terrestrial vs. lunar is cheaper. Delta-v values are approximate.

Fig. 6 shows the results after changing to SEP to deliver the lunar product below LLO. The slope of prices changes below LLO due to the higher $I_{SP}$ despite the increasing finance cost with transit time. With this change, lunar propellant has an absolute advantage in GTO by year 3 and to LEO by Year 15.

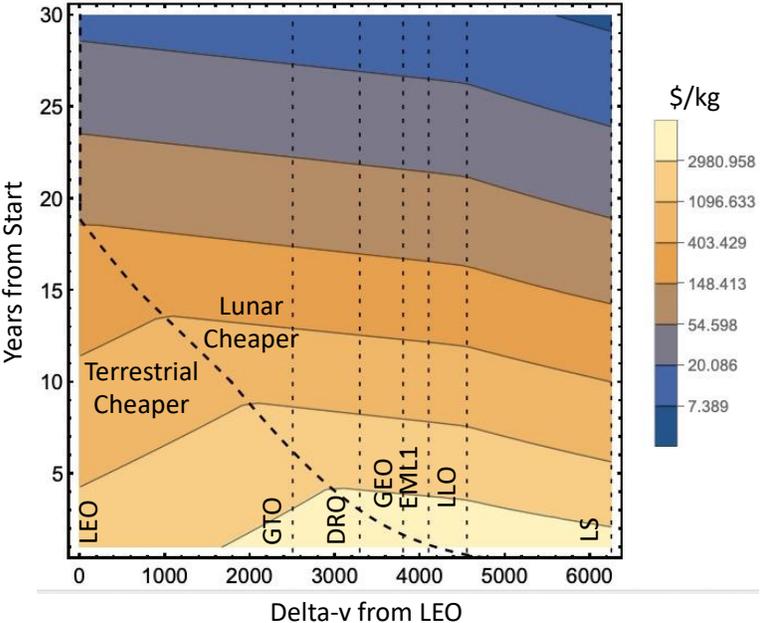

**Figure 6.** Same as figure 5 (optimistic market) except it uses SEP to deliver lunar propellant between LLO and LEO.

Fig. 7 shows the proportions of cost due to labor, capital, and finance. Over time, capital becomes an increasing share of the costs relative to labor. Finance causes 82% of costs in year 1 dropping to 73% by year 30. This suggests a large role that government could play to improve the economics of this sector in its startup years, by creating a public-private partnership (PPP) to lower investor risk, which lowers the acceptable rate of return and thus dramatically reduces cost. Eventually proof of successful operation will lower the investor risk, so this PPP is needed only at the start.

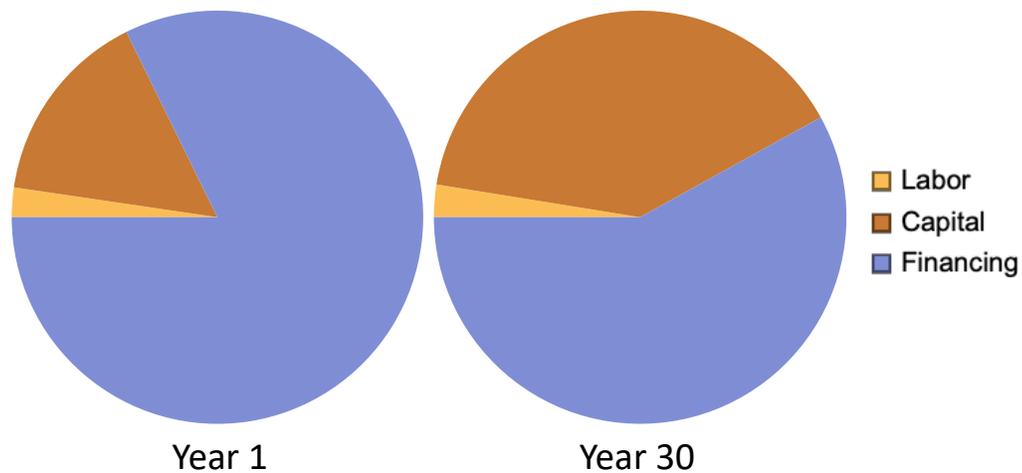

**Figure 7.** Proportion of costs due to labor, capital, and financing for years 1 and 30.

## *5.2. Baseline Case, Moderate and Pessimistic Market Scenarios*

The baseline case was also evaluated with moderate and pessimistic markets by setting $D_{30}$ to 10% and 1%, respectively, of the optimistic value, which corresponds to much lower market elasticities, so these are indeed pessimistic. The launch cost curve was recalculated per Eq. 17-18, and the market model was recalculated per Eq. 19. Because the lesser market reduces EOS and learning curve both for terrestrial propellant launch cost and for lunar propellant, the changes to the competitiveness are small, as shown in Fig. 8. Table 1 shows the years of operation until lunar achieves absolute advantage at various locations in cislunar space. Absolute advantage reaches to GTO in almost the same time in each case. Only LEO is noticeably affected: each order of magnitude reduction in the market delays absolute advantage in LEO by about 2 years. This demonstrates that the conclusions of this study do not depend strongly on the size of the market. Taken together, figs. 6 and 8 indicate that the economic viability of lunar resources is not strongly dependent on launch costs or market size, disproving the claim by skeptics that low launch cost will drive space resources out of business.

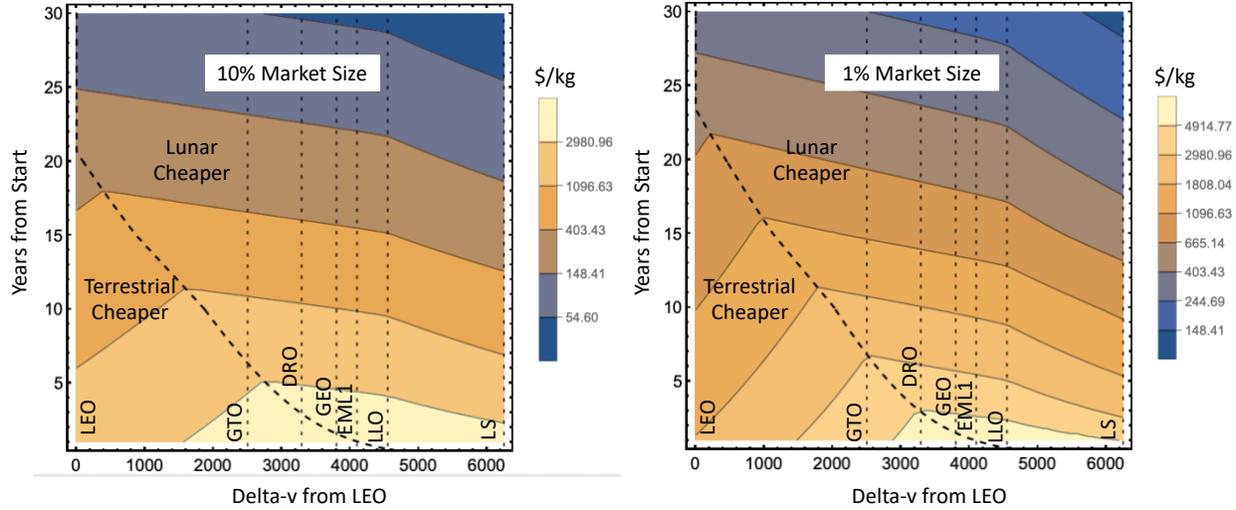

**Figure 8.** Same as figure 6 except for a moderate market 10% of baseline (Left) and a pessimistic market 1% of baseline (Right).

**Table 1.** Years of operation until lunar propellant gains absolute advantage in various locations in cislunar space. Calculated for baseline technology parameters in three market size scenarios.

|      | Optimistic (100%) | Moderate (10%) | Pessimistic (1%) |
|------|-------------------|----------------|------------------|
| LS   | 1                 | 1              | 1                |
| LLO  | 1                 | 1              | 1                |
| EML1 | 1                 | 1              | 1                |
| GEO  | 2                 | 2              | 2                |
| DRO  | 3                 | 3              | 3                |
| GTO  | 6                 | 7              | 7                |
| LEO  | 19                | 21             | 23               |

## 5.3. *Sensitivity to Other Economics Parameters*

The baseline case using SEP in the optimistic market scenario was re-run with different values of economic parameters to test their sensitivity. The results are shown in Fig. 9. The Supplementary Material shows the long-run trends in each of the non-dimensional cost parameters of Eqs. 8 and 9. Ending the firm-level EOS (but not the supply chain EOS) at either production rate $X_{max} = 10$ t/day or $X_{max} = 20$ t/day caused the cost ratio to become flatter, but it happened late enough that the year 1 cost ratio was already at the threshold for advantage in GTO. Eliminating the SOE in the supply chain by setting $\beta = 0$ or increasing SOE by setting $\beta = 1$ (i.e., other in-space industries grow to match propellant mining) had negligible effect. Reducing the EOS by setting $a = 0.8$ caused a modest slowing of the curve. Varying the learning curve parameter between

$b = 0.70$ and $0.80$ had a modest effect. It is difficult to imagine lunar industry having a value $b > 0.80$.

The implication of this plot is that lunar propellant will continuously gain large additional advantage over terrestrial propellant during the first 10 to 30 years of operation, and possibly longer. Terrestrial propellant price is exponentially dropping during this period, but lunar propellant drops faster. The timescale of these curves is short enough that the change is relevant for planning both national policy and space agency strategy, i.e., it would be a mistake to set policy based upon the short-run costs predicted by the TEAs, alone. The tightness of the bundle of curves in Fig. 9 suggests that this is a firm conclusion.

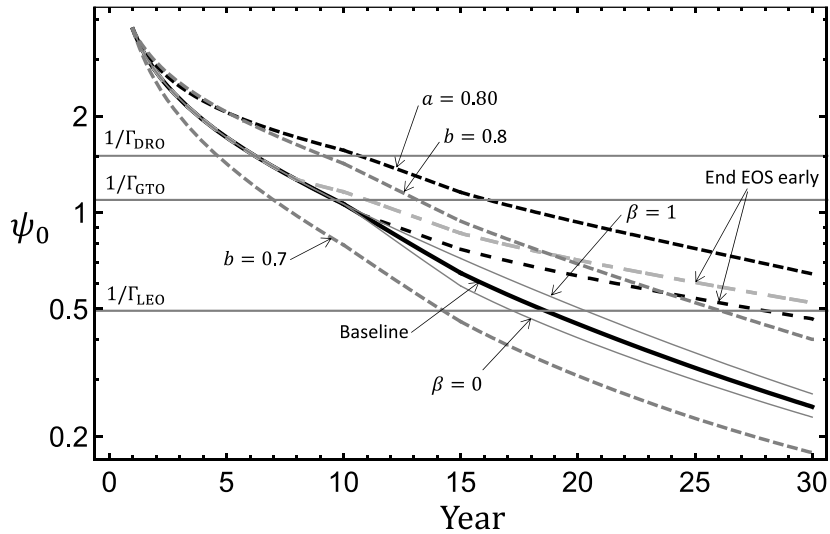

**Figure 9.** Pre-delivery cost ratio of lunar vs. terrestrial propellant for varied economic parameters. When the curves cross below the thresholds $1/\Gamma_{DRO}$, $1/\Gamma_{GTO}$, or $1/\Gamma_{LEO}$, lunar propellant has achieved an absolute advantage over terrestrial propellant in that location.

### 5.4. *Sensitivity to Technology Parameters*

The same scenario was re-run varying the technology parameters one at a time to identify sensitivities as shown in Fig. 10. $\phi$ is varied by changing the mass of product $M_{p,LS}$ in the numerator while leaving the mass of capital $M_K$ in the denominator unchanged. The effects of changing $\phi$ and $M_K$ are nearly inverse but not precisely so because changing $M_{p,LS}$ also affects the specific operations cost while $M_K$ does not, and changing $M_K$ affects capital transportation costs while $M_{p,LS}$ does not. Changing $M_K$, $M_{p,LS}$ and $\zeta$ (i.e., both $\zeta_D$ and $\zeta_F$) have a nearly identical direct impact on cost in Year 1. Entrepreneurs may lower the cost of lunar propellant directly by making improvements in any of the three parameters. Over time, $\zeta$ has diminishing effect while $M_K$ has increasing effect due to EOS reducing the cost of capital while $G$ stays unchanged in physics so transportation becomes relatively more important. The Supplementary Material shows the behavior of the individual non-dimensional cost parameters of Eqs. 8 and 9.

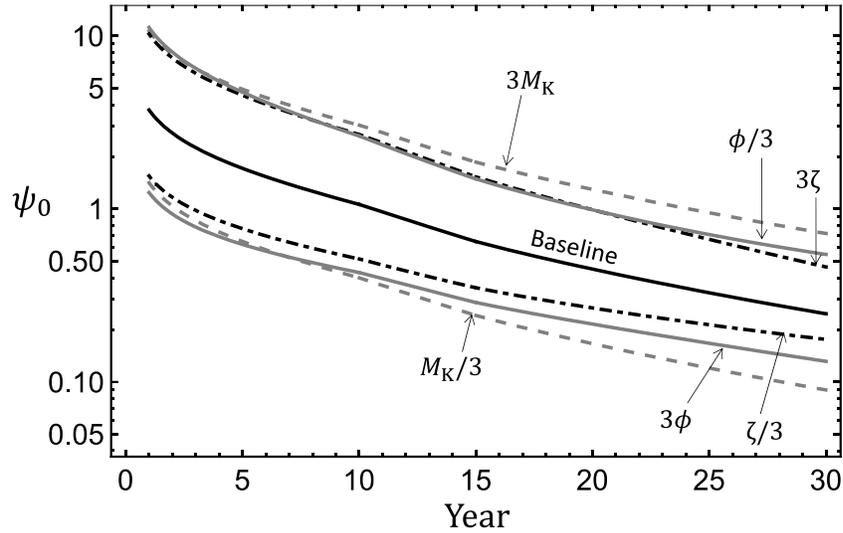

**Figure 10.** Pre-delivery cost ratio of lunar vs. terrestrial propellant for varied technology parameters

Maximizing $\phi$ appears to be the dominant strategy for lowering the cost of lunar propellant – i.e., maximizing the mass of product per mass of capital. The values of $\phi$ estimated by the extant TEAs are shown in Table 2 along with the value used in the baseline model, which was chosen to be mid-range. Because of the role of $\phi$, the TEAs for tent sublimation are more optimistic than this baseline while the TEAs for strip mining are less optimistic.

**Table 2.** Values of $\phi$ per the extant studies

| Mining Method | Tent Sublimation Technology | Borehole Sublimation Technology | Strip-Mining Technology | | | | Beneficiation Technology | Baseline Model |
|---|---|---|---|---|---|---|---|---|
| Study | K | S | P | CD | J | B | P | M | |
| $\phi$ | 442 | 534 | 16.1 | 26.5 | 22.2 | 43.4 | 3.7 | 36.5 | 167 |
| Abbreviations for studies: K = Kornuta et al.[1], S = Sowers [4], P = Pelech et al. [7], CD = Charania and DePascuale [5], J = Jones et al. [6], B = Bennett et al. [2]; M = Metzger et al. [3]. | | | | | | | | | |

The parameter elasticities on the cost ratio $\psi_0$ were tested by running the model with 1% increments of a parameter and are shown in Table 3. The elasticity for $M_{p,LS}$ can be interpreted as the (negative of) elasticity for $\phi$. It is nearly equal to (negative) unity since it divides directly into cost and has no offsetting effect on the competing terrestrial propellant costs. $M_K$ also affects $\phi$ but it is entangled with hardware fabrication cost and transportation cost so elasticity for $M_K$ is not identical to elasticity for $\phi$. Elasticity with $I_{SP}$ (for capital delivery) exceeds (negative) unity when transportation costs dominate because of the exponential dependence in the rocket equation. The cost ratio is nearly inelastic on launch cost $L_0$ when $G \gg x$, since it affects both terrestrial and lunar propellant costs equally. It is nearly (negative) unity when capital acquisition costs dominate. Lunar propellant can therefore be insulated from decreasing launch costs by achieving $x < G$ as a capital design goal.

**Table 3.** Parameter Elasticities on Cost as a Function of the $G/x$ Ratio

|  | $M_{p,LS}$ | $M_K$ | $\zeta$ | $G$ | IMF | $I_{SP}$ | $L_0$ |
|---|---|---|---|---|---|---|---|
| $G/x = 0.02$ | -0.990 | 0.983 | 0.973 | 0.011 | 0.005 | -0.023 | -0.947 |
| $G/x = 1$ | -0.990 | 0.710 | 0.432 | 0.277 | 0.120 | -0.614 | -0.692 |
| $G/x = 50$ | -0.990 | 0.982 | 0.024 | 0.958 | 0.437 | -2.116 | -.0.040 |

## 6. **Assessing the TEAs**

Four studies were optimistic about the competitiveness of lunar propellant beyond LS or LLO: Kornuta et al. [1], Bennett et al. [2], Metzger et al. [3], and Sowers [4]. Three studies were pessimistic: Charania and DePascuale [5], Jones et al. [6], and Pelech et al. [7]. Here we re-assess these studies considering especially how they handled $G$ and $\phi$.

### *6.1. Charania and DePascuale*

Fig. 11 considers five variations of the Charania and DePascuale study [5], hereafter CD. Curve (1) replicates CD normalized by launch cost, $L_p(t)$. CD only predicted short run costs, so the curve for subsequent years was calculated using the same economic scaling parameters as the baseline model. CD assumed high $G = 64.9$ (per current pricing) due to the use of government-built (non-commercial) rockets for capital transport. Even with EOS/SOE and learning curve, since $G$ is high the reduction in $x$ has diminishing returns and the slope in the curve is small. Because of the technology and transportation assumptions in CD, the curve is two orders of magnitude too high to be economic.

CD was written before the LCROSS mission that proved lunar ice concentrations at 5% and higher exist [94], so CD assumed 1% ice concentration. Current thinking in the lunar mining community is that nobody would pursue a deposit with such low concentration. Curve (2) makes only one change: it assumes a prospecting campaign has identified a better ore body that obtains 5% yield so $\phi$ is improved by a factor of 5.

Curve (3) makes three more changes. First, it uses commercial launch of the capital assets, so $G$ is reduced from 64.9 to 8. This reduces the cost of propellant at LS by 59% in year 1 which grows to extreme cost savings by year 20 and beyond since $G$ is now lower than $x$ and the returns from EOS and learning curve are not diminished. Second, it implemented optimization of reliability, which provides another 10% reduction in cost at LS. Third, it implements SEP for delivery of the product from LLO toward LEO. The longer delivery time increased finance costs but also raised the thresholds $1/\Gamma_X$ where lunar propellant gains the absolute advantage. The three thresholds shown on the plot are therefore only applicable to Curves (3) through (5) that use SEP. This curve gains absolute advantage in DRO by year 15 and in GTO by year 18, expanding the business case beyond customers near the Moon.

Curve (4) recognizes that NASA is currently working toward a lunar pilot plant for ice mining and propellant production, so the buildup time of 8 years in CD may be reduced to 4 years and 50% of the development cost $\zeta_D$ (but not the fabrication cost $\zeta_F$) of equipment may be covered by that government-funded effort. Curve (5) now gains absolute advantage for lunar propellant in GTO by year 13. This is still a long time but the high discount rate of 27.2% is commensurate.

Curve (5) assumes a PPP is established and the discount rate is lowered to 12%. Now, it gains absolute advantage in GTO by year 8. The CD study becomes optimistic for lunar resources with these changes.

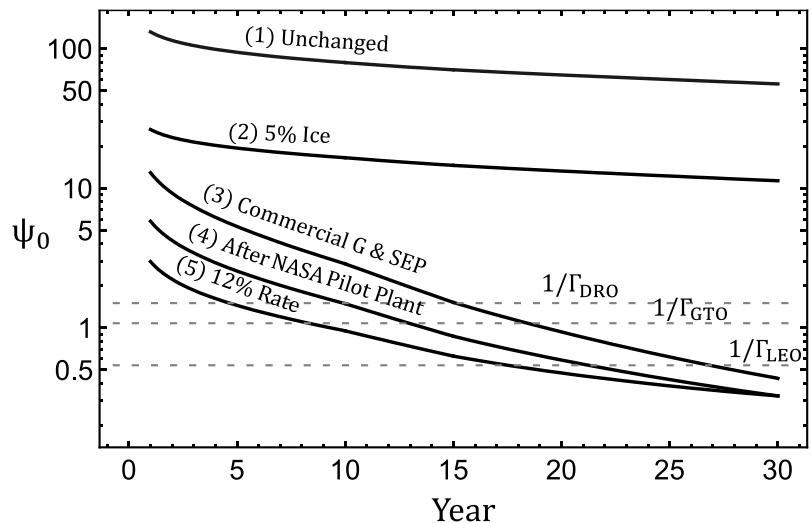

**Figure 11.** Pre-delivery cost ratio for CD with four improvements. The dotted lines are thresholds for cases with SEP (cases 3 to 5). When crossed, lunar propellant has gained absolute advantage over terrestrial at the corresponding location (DRO, GTO, or LEO).

### 6.2. *Jones et al. and Bennet et al.*

The study by Jones, et al. [6], hereafter J, evaluated a strip mining approach and predicted a low value of $\phi = 22.2$. Bennett et al. [2], hereafter B, assessed the modeling by J and showed that different design choices can increase the economies of scale for the reactors, resulting in $\phi = 43.4$. J used the Space Launch System (SLS) for capital transport to LS. At the time, SLS was a good alternative with $G = 5.4$. Since commercial Heavy Lift Launch Vehicle (HLLV) prices have already dropped but SLS costs have not dropped, SLS now yields $G = 41.8$. This would dominate the capital costs and ensure future reductions in its fabrication cost would have negligible return. J neglected finance and operations costs, probably because it was assessing a government-led effort without full-cost accounting. If they had been included, J would have an even more pessimistic prediction. Fig. 12 shows the resulting cost curves for J and B using the same capital transport as the baseline model instead of SLS, with reliability optimized, and finance and operations cost added using the parameters in Table A-1. The results are not as optimistic as predicted by B due to the added finance and operations costs. It does not gain absolute advantage in DRO until year 16 or in GTO until year 20. However, it indicates lunar resources are competitive for refueling at LS from the start. Sowers [4] explained that the method J used to approximate $\phi$ (which B followed) was methodologically flawed. It was based on an inappropriate analogy to Molten Regolith Electrolysis (MRE), which is unlike ice extraction. Even with the economies of scale corrected by B, this analogy cannot prove the strip mining technology to be either economic or uneconomic. In contrast, CD was based on a detailed design study by CSP, Japan, Inc., and Shimizu Corporation. The details of that study are not published but are described at a top-level by Charania and Kanamori [95]. A detailed, published TEA for the strip mining technology is still needed.

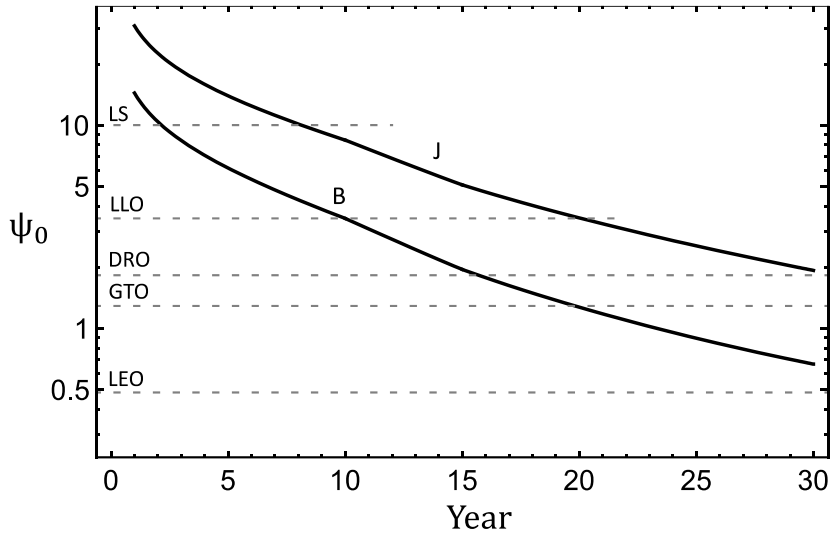

**Figure 12.** Pre-delivery cost ratio for J and B using SEP. The dotted lines are the thresholds for absolute advantage at each location, $1/\Gamma_X$.

### 6.3. *Pelech et al.*

Pelech et al. [7], henceforth P, assessed two technologies for lunar ice mining. The first was strip mining the icy regolith then thermally separating ice from the excavated soil by sublimation, similar to CD, J, and B. The second was to drill boreholes then sublimate the ice in situ by injecting energy (e.g., microwave) down the boreholes using battery-stored energy onboard a rover. P did not assess cost but only the "propellant payback ratio" $PPR = G^{-1} \phi \, \Gamma_X^{-1}$, the inverse of the (launch-normalized, product-specific) capital delivery cost. They argued the PPR is too low in these two technologies so lunar propellant will not be economic in the near term.

PPR depends on the transportation architecture for G and $\Gamma_X$ and on the design of the capital for $\phi = M_{p,LS}/M_K$. P assumed capital would be delivered by the Falcon Heavy rocket for $G = 7.5$, which is competitive. For lunar propellant delivery P used an RLL with $IMF = 0.83$, which is unrealistically high. This created a 50% inefficiency in delivering the product to LEO, contributing somewhat to the low PPR. More importantly, P estimated that $\phi$ will be far too low in both technologies.

For the strip mining technology P roughly estimated $\phi \sim 3.7$. CD, J, and B predicted $\phi$ an order of magnitude larger for similar technology. The question becomes which of these studies has the more reliable estimate. None of the studies published enough detail to answer this. Pelech (personal communication, T.M. Pelech, Aug. 16, 2022) stated that the technology details in P were relatively less important because the study's intent was to demonstrate the usefulness of the PPR, so significant changes in the estimated $M_K$ are possible. In P, the estimate of the bucket wheel excavator mass was based upon an analogous terrestrial excavator after scaling the force equations, but space technology uses materials and designs to achieve lower mass without compromising the forces, so a terrestrial analogy should produce too high an estimate of $M_K$. NASA showed that using carbon fiber and aluminum instead of steel reduces the mass of a large excavator blade by 70% without sacrificing strength [96], and that reaction forces for digging buckets can be reduced via percussion in low gravity to avoid using more mass [97,98]. These technologies are not used in terrestrial applications because transportation of capital is cheap, so they seek to minimize the cost of capital rather than its mass. Although P did not assess cost, these mass-savings are typically reflected in the higher fabrication cost of spaceflight hardware. Since we are already paying for them, it is appropriate to include them in the mass estimate. It is impossible to explain the full difference in $\phi$ between these studies without more detail on how the costs were estimated, so the economic viability of strip mining lunar ice is still undecided although the CD study with changes discussed above favors its viability.

For the borehole sublimation technology, the values in P indicate a much better $\phi \sim 16.1$. If $M_K$ was overestimated by a factor of 3 then this puts this technology in the economic range. It appears P determined the mass of the drilling rover by the mass needed for weight on bit in lunar gravity, rather than the mass needed to fabricate the chassis and subsystems in a bottom-up assessment. Any shortage in the weight on bit can be made up by filling regolith in a bin on the rover or by anchoring [99], and drilling methods have been developed to reduce the necessary weight on bit in low gravity environments [100]. Delivering a more massive rover to provide weight on bit is therefore not needed and contributes to a lower $\phi$. The estimate of the mass of the downhole energy rover starts with a rough estimate that batteries constitute 50% of the mass, followed by a calculation of the mass of batteries. Regenerable fuel cells would require only half the mass of batteries, and the 50% figure may not be accurate for spaceflight construction. Without more detail, we cannot say whether the borehole sublimation is economic, but it appears to be in the range of becoming economic. The focus is clearly on the technology design to improve $\phi$.

### 6.4. *Kornuta et al. and Sowers*

Sowers innovated the tent sublimation technology with the goal of vastly increasing $\phi$. It became the subject of the United Launch Alliance's study authored by Kornuta et al. [1], henceforth K. Another analysis was authored by Sowers [4], henceforth S. K predicted $\phi = 441.8$ and S predicted $\phi = 534$. This will clearly be economic and will outcompete other lunar mining methods if the technology works as expected. Bennett et al. [13] analyzed a business model based on K including a propellant delivery architecture showing it to be profitable. Recent experimental work by Sowers [101] and Purrington et al. [102] indicates that the technology will work. S based the demand estimate on a portfolio of applications in cislunar space, between 40% and 100% for use in LEO per the three scenarios.

S estimates an Internal Rate of Return (IRR) of 8.84% in a fully commercial venture, which is much less than the 21.7% that CD thought necessary to attract investors. S describes how a public-private partnership can offset development costs to raise the IRR while also making a lower rate of return acceptable to investors since the government bears a portion of the risk. S did not model how the costs may evolve in subsequent years and for conservatism ignored economies of scale (George Sowers, personal communication, May 15, 2022), but the modeling of cost evolution here shows that even with the 21.7% rate it may achieve absolute advantage to LEO by Year 5. Based on the results of K and S, there is a clear business case so navigating the startup financing is the remaining challenge.

*6.5. Metzger et al.*

The study by Metzger et al. [3], henceforth M, had the goal of solving the startup finance problem. Two prominent space mining companies (asteroid mining companies) had raised seed and Series A investment rounds but were unable to secure their Series B rounds [103]. M approached this by developing a "Minimum Viable Product" technology that should produce net revenue immediately for the minimum $M_K$ and hence the minimum initial investment, although it sacrifices $\phi$. The mining company would quickly operate in the lunar environment to mature the technology and buy-down risk while making revenue to offset costs. The most important product would not be propellant but investor optimism. Successful Series A and subsequent rounds would enable scale-up of the operation, including transition to larger-scale mining methods for greater $\phi$. The technology that was innovated for this [104] was based on a geological argument that the ice should be in granular form that is highly excavatable, so a low-energy beneficiation process can reduce the excavated mass by 95% prior to hauling ore into a sunlit (energy-rich) location for processing. This eliminates the need to beam energy into the permanently shadowed regions where the ore bodies are found, dramatically reducing the mass of energy infrastructure. M estimated the resulting $M_K$ to be an order of magnitude lower than other studies, so startup costs are lowered by an order of magnitude, yet $\phi = 36.5$ is still high enough to gain absolute advantage at least to GTO as shown below. It would produce enough propellant to boost four satellites from GTO to GEO annually. Ground truth of the lunar ice is needed to validate the geological arguments in M. NASA is currently developing a mission to obtain ground truth on the ice [104].

*6.6. Other Technological Approaches*

There are other technologies in development that I did not discuss only because I do not have access to TEAs on them. Ethridge and Kaukler [105–107] and Ethridge [108] developed down-borehole heating using microwave or radio frequency energy. Honeybee Robotics developed the Mobile In-Situ Water Extractor (MISWE) and Planetary Volatile Extractor (PVEx) technologies [109,110], both of which drill a borehole, the former bringing the cuttings to the surface via an auger for thermal volatile extraction, the latter heating the icy regolith within the borehole for thermal extraction. Sercel [111] and Sercel et al. [112] innovated a multi-wavelength, ground penetrating beam method powered by tall tensegrity towers to capture the solar energy above the Moon's permanently shadowed craters. Kuhns et al. [113] innovated "rocket mining" whereby a rocket thruster under a dome disrupts and heats the soil rapidly to liberate and capture volatiles. Austin et al. [14] analyzed architectures for robotic lunar ice mining based on three alternative approaches: mechanical strip mining, pneumatic strip mining with pneumatic beneficiation, and the Honeybee Robotics MSWE technology. There are several companies working on lunar ice mining so there may be other technologies or architectures of which I am unaware.

*6.7. Summary of Prior Studies*

The parameter values in Table A-1 were provided by the different extant studies. Some studies did not include or report various parameters, so each study was back-filled as explained in the notes of Table A-1. Optimization of reliability was added onto each case, and the same long-run economic scaling relationships were applied to each. To compare technologies on an equal basis, all studies were modified to use the same transportation parameters discussed above. K and S both considered transportation to be the business of a different firm than the lunar mining company, so they did not include development of transportation capital but here they have been included. Fig. 13 shows the long-run average cost for the studies. CD5 is the modified version of CD from Curve 5 of Fig. 11.

The curves fall into groups based on technology. This justifies the statement in the introduction that an individual TEA cannot disprove the viability of ice mining as a sector but only of a particular technology (and then only in a specific form since future innovations may improve the technology). The clustering also shows the key role of $\phi$ in determining economic viability. All models except J and B indicate that an absolute advantage is gained in GTO no later than year 12. Both tent sublimation studies predict economic viability in LEO from year 5.

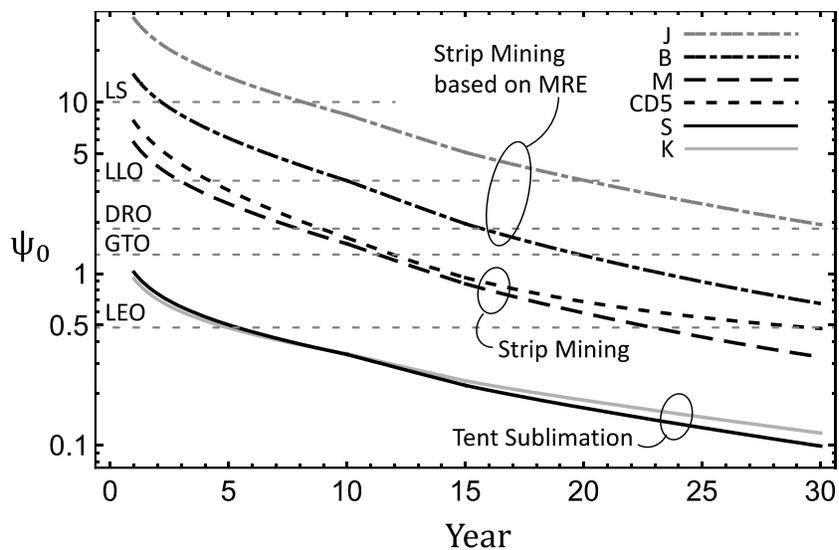

**Figure 13.** Pre-delivery cost ratios for six studies with SEP using 27.2% discount rate. CD5 = modified CD per curve 5 on Fig. 11. See text for study identifiers.

$1/\Gamma_{DRO}$, $1/\Gamma_{GTO}$, or $1/\Gamma_{LEO}$, lunar propellant has achieved an absolute advantage over terrestrial propellant in that orbit.

## 7. Sensitivity to Discount Rate and Market Size

Figure 15 shows the same six studies as Figure 13 but it assumes they are implemented in a Public-Private Partnership so the discount rate is 12% (constant over 30 years). Now, all studies except J and B predict absolute advantage in GTO by no later than year 10.

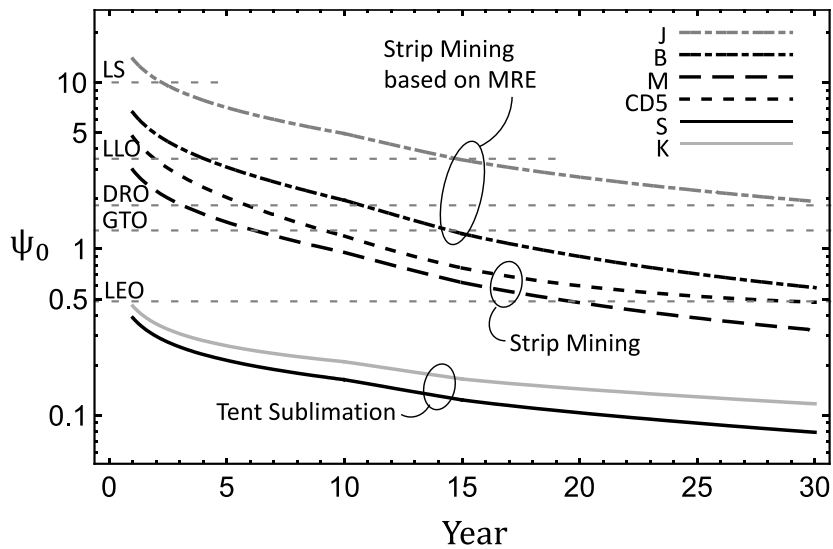

**Figure 15.** Pre-delivery cost ratios for six studies using 12% discount rate. CD5 = modified CD per curve 5 on Fig. 11. See text for study identifiers.

The two main technological approaches, strip mining and tent sublimation, were tested against the optimistic, moderate, and pessimistic market cases to see if either technology is more sensitive to the scale of the market. As shown in Fig. 14, the scale of the market has negligible effect until about 10 years of operation and only a minor effect after that. Tent sublimation still has absolute advantage all the way to LEO from year 1. Strip mining is delayed for gaining absolute advantage in LEO in the more pessimistic market but still has a business case down to GTO without additional delay.

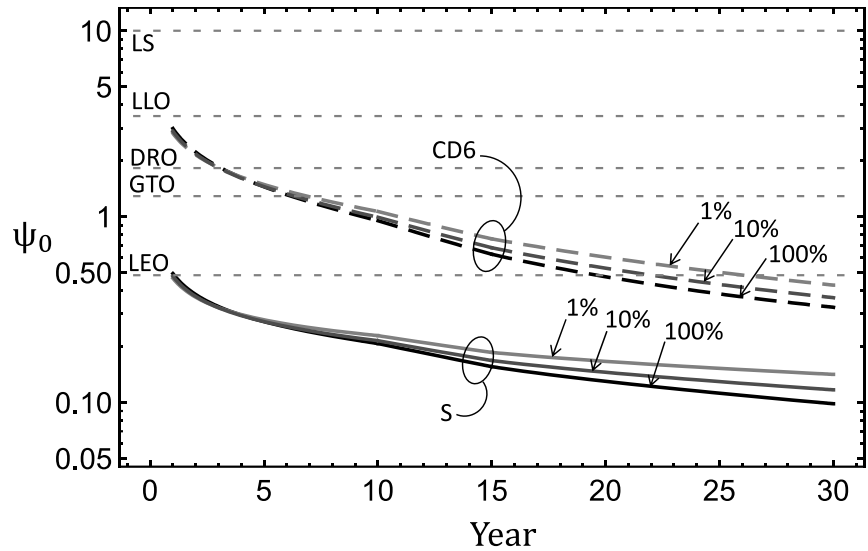

**Figure 14.** Pre-delivery cost ratios for CD6 strip mining (dashed curves) and S tent sublimation (solid curves) each at 12% discount rate. Black: baseline market model. Gray: market model reduced to 1% of baseline.

## 8. Sensitivity to Reliability

To test sensitivity to reliability, additional cases were modeled for both strip mining and tent sublimation while varying $R_0$. A brief explanation is needed to explain the difference between varying $R_0$ and varying $R$. It is a difference between innovating new technology concepts during Technology Readiness Level (TRL) 1 though 6 versus execution of established technology concepts during TRL 7 through 9 [114,115]. For the latter, if we build flight hardware using normative spaceflight parts, materials, design features, fabrication processes, and preflight testing, then when operating on the Moon the actual reliability of the hardware $R$ (or the "as-built reliability") will equal the baseline $R_0$ by definition. Building the flight hardware with better or worse parts and processes and with more or less testing than the norms for spaceflight will yield higher or lower $R$, respectively, but will not change $R_0$. These decisions are what drive the exponential cost of reliability as described by Mettas [33].

The baseline reliability $R_0$ on the other hand can be improved through innovation and maturation of new technological *concepts* along the TRL scale from 1 to 6. To improve $R_0$ a new technology concept must address one or more of the following described by MIL-HDBK-338B [36]: (1) reduce exposure to harsh environmental and operating conditions; (2) reduce operating time; (3) simplify the technology so it is less "intricate" (e.g., fewer parts); and (4) mature the technology to identify and address failure modes. For a rover that must operate in harsh dust we might improve $R$ by using more powerful motors and higher-torque gear assemblies that can continue to turn despite the friction of dust accumulated in rotating wheel joints, or we might improve $R_0$ by inventing mobility system that uses reciprocating rather than rotating joints so they can be booted, preventing dust from entering the joints in the first place.

All other factors being equal, improving $R$ increases the cost of the hardware but improving $R_0$ decreases the cost. After the new technology concept with higher $R_0$ is matured to TRL-6 and adopted for in-space use, the team that builds the flight hardware from TRL-6 to 9 can achieve the targeted value of $R$ without using the more expensive and massive motors or other parts, or more extensive testing, etc. Therefore, varying $R_0$ in the model represents technological innovation and maturation of new technologies before incorporation into flight-like builds. Since so little work has been done in lunar mining, there is room for much innovation to lower $R_0$, but since we have no data from actual lunar mining there is also uncertainty in what value of $R_0$ to use for the current state of the art, hence the need to vary it here.

The model assumes a value for the baseline $R_0$ then optimizes $R$ for minimum overall cost using the exponential cost of reliability relationship, Eq. 11. The results are shown in Fig. 16 where there is a set of curves for CD5 (representing strip mining) and a set of curves for S (representing tent sublimation). If $R_0 = 0.55$, 0.70, or 0.85 can be achieved, then strip mining would gain absolute advantage in GTO by year 9, 7, or 5 respectively and gain it in LEO by year 24, 21, or 17 respectively. For tent sublimation, all cases have absolute advantage in GTO by year 1. Even if $R_0$ for tent sublimation were as low as 0.10, it would have absolute advantage in LEO by year 5. $R_0$ was incremented in uniform $\Delta R_0 = 0.15$ steps within each set of curves but the resulting cost differences between the curves are not uniform increments. This is because the cost of reliability while optimizing $R$ is exponential, not linear. As $R_0$ achieves increasingly higher values through technology innovation, the reductions in cost become increasingly greater. This emphasizes the important role of innovation.

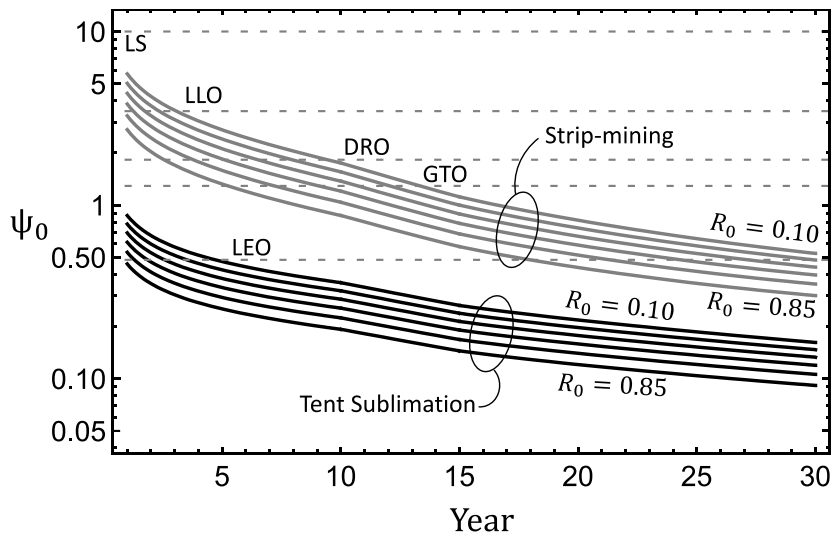

**Figure 16.** Pre-delivery cost ratios for CD5 strip mining (gray curves) and S tent sublimation (black curves) assuming baseline reliabilities $R_0 = 0.10$ (top curve in each set), 0.25, 0.40, 0.55, 0.70, and 0.85 (bottom curve in each set).

## 9. Discussion

*9.1. Concerns of Inadequate Reliability*

Several members of the aerospace community expressed concern that lunar propellant may not become competitive because of technological and business pragmatics. Several mentioned their concerns with reliability, either that it will be too hard to achieve adequate reliability in the harsh lunar environment, or that firms will not be able to lower costs by lowering reliability to some intermediate level between "too reliable and too costly" and "not reliable at all." Ultimately, the only way to remove these concerns is to build and test hardware, advancing it to TRL 6 where the tests include the relevant environment, and then to operate the system on the Moon. That work is expensive so decision-makers will want to confidence to proceed. This modeling suggests we have adequate confidence for that next step. We may note that NASA is already driving technology maturation for lunar ice mining through contract and grant solicitations and through the Break the Ice Lunar Challenge, part of the NASA Centennial Challenges program [116].

The most serious challenge to reliability may be the abrasive lunar dust. Some of the technologies, like tent sublimation, do not strip mine so they have much less environmental stress and it may be easier to achieve the reliability targets. System-average reliability can be improved by making assets more modularized for telerobotic replacement of smaller units. It should be remembered that most of the mass of assets will have higher reliability than the norm, because it is usually a small fraction of the mass that fails often. This improves the average reliability. Open-ended innovation can improve $R_0$ to reduce costs non-linearly as demonstrated above. Even if $R_0 = 0.10$, which is unreasonably low, at least one technology is still economic per Fig. 16.

*9.2. Concerns of Inadequate Robotic Autonomy*

Some in the community point to the degree of robotic autonomy as another cause for doubt. The economic analysis here assumes robotics will operate without humans present. Including humans greatly increases cost but also greatly increases capability to troubleshoot and repair hardware and to study improvements in the technology. There may be broader business cases than discussed here that include humans for greater revenue-generating scope than just propellant mining, which would offset the costs. For one example, there are non-commercial reasons for national space agencies to have humans on the Moon doing lunar science and technology maturation in support of national objectives. As transportation costs to the lunar surface drop, the cost of humans on the Moon will also drop.

On the other hand, many in the community are convinced we can build a fully robotic operation with no humans on-site or only minimal human visits. One source of confidence is the demonstration of progress by the NASA Lunabotics competition [117–120] and the NASA Break the Ice Challenge [116]. In the first few years of Lunabotics, university student clubs produced robots that were barely functional in dusty lunar-like regolith. After 10 years of their learning curve, with knowledge passed down informally within and between the clubs, they were building highly capable lunar mining robots that were fully autonomous and robust in the harsh regolith. The competition judges studied the robots and kept records of their improvements over the years but never shared this data with the students so as not to disturb the competition. In the first year, almost every robot was immediately stuck in the regolith. After a few years, some teams had discovered how to make robots nearly immune to getting stuck. Some robots used vision systems with machine learning to identify and navigate around terrain hazards, automatic feedback to adjust motor torques and driving speeds, localization and area navigation to find and then drive to and from the mining zones, feedback controls in the mining systems to adjust for changes in digging conditions, and docking sensors to align the robot with the simulated chemical processing plant to offload the ore. Some teams implemented multiple layers of navigation for redundancy. Robots could drive back and forth in repeated mining cycles with no human intervention. The teams engaged in continuous innovation in all the robot's subsystems year-by-year. The clubs accomplished this using off-the-shelf components such as mobile phones and gaming consoles, and with budgets on the order of $5,000 to $10,000 USD and a labor force that worked only part-time in the evenings without pay. It seems unlikely that full-time professional technologists with corporate support should be unable to make similar progress using space-qualified components and a much larger budget.

At an early stage, autonomy can manage discrete tasks such as driving around obstacles and moving the digging bucket to take an optimized bite of soil – tasks that enable teleoperation despite the 3-second round-trip communications latency between the Earth and Moon. Part of the expectation for a long-run learning curve in this study is that autonomy will continue to improve over the next thirty years so greater autonomy can be phased-in smoothly across the various tasks and integrated operations, asymptotically approaching full autonomy. Autonomy is already advancing rapidly in terrestrial mining, construction, and manufacturing so this expectation is well supported. When human intervention is needed, the shortness of the communications latency allows much faster exchanges than with Mars rovers, making loosely supervised autonomy feasible with fewer teleoperators per robot standing by for support year-by-year.

*9.3. Concerns of Inadequate Cost Reductions*

There is also a question whether a learning curve or economies of scale can really reduce costs in lunar industry since it is unlike terrestrial industry in some ways. Can the processes really be simplified and improved to reduce costs in every increment? I argue there is no reason to doubt it, because lunar industry will be identical to terrestrial industry in all the ways that matter. Capital will still be made on Earth (at least at first), benefitting from improvements in the terrestrial factories and supply chain just like any other industry. Next generation hardware will be redesigned for simplicity and efficiency, as usual. Operations on the Moon can be simplified as we learn how to position equipment around an ore body and schedule tasks efficiently, just like any terrestrial operation. The payload farings on future rockets like the SpaceX Starship will be so large that it seems impossible we should need larger diameter capital assets for the projected throughput of product. Even if we do, that has already been included conservatively in the model through $X_{max}$ in Eq. 16. The throughput rate of water cleanup systems and electrolysis at $X_{max} = 10$ t/day allows 2.4 hours for every cubic meter of water that is processed, and that needs only a tiny pipe, orders of magnitude smaller than a payload faring, so it seems unlikely that EOS limitations really exist on these scales. The only ways that lunar industry will differ from terrestrial industry are (1) the transportation costs, which have been included in the model, (2) the harshness of the environment, which has been modeled in the reliability equations, and

(3) the lack of human access to repair the capital in its operational location. Further technology maturation is needed to address these last two challenges, but they provide more opportunity for learning curve, not less.

*9.4. Possible Extensions of the Analysis*

A similar analysis could be done for transporting capital to Mars to support operations on that planet. In that case, there are more options to decrease $G$ because propellant manufacturing can take place on Mars, Phobos, and the Moon to create multiple refueling stations in the vicinity of both Earth and Mars to break up the exponentiality of the rocket equation. A large fraction of the technologies and infrastructure for lunar propellant mining can be adapted to Phobos and Mars, so developing lunar propellant mining and manufacturing at the present time on the Moon will have long-term benefit for Mars and beyond.

Another way to reduce the cost of transporting capital is to get as much of its mass as possible in situ on the Moon or Mars rather than transporting it from Earth. Here, we considered metal manufacturing only for its effects lowering supply chain costs via economies of scope. Using the metals to fabricate structural elements and other parts of the lunar propellant firm's capital may bring an additional reduction of costs that bypasses the limitations on $G$. Future analyses should modify the equations to include that dynamic.

## 10. Summary

This analysis has shown that the profitability of lunar propellant is possible in physics and plausible in economics as a function of hardware performance parameters that are subject to open-ended improvement. It depends mostly on reasonable $G$ and high $\phi$. The former has a limit in physics, but the latter does not. The threshold of absolute advantage in this modeling is $\phi \gtrsim 35$. The tent sublimation method was innovated to obtain higher $\phi$ and its values are predicted to be an order of magnitude better than the threshold of absolute advantage. The strip mining method is closer to the threshold. No adequately detailed, publicly available TEA exists to parse the narrower margin for strip mining. However, when the EOS and learning curve are included, strip mining gains absolute advantage over terrestrial propellant in GTO within just a few years of operation. Furthermore, any technology that is close to the threshold like strip mining can become profitable simply with more innovation. Lowering $M_K$ and increasing $M_{p,LS}$ are analogous to the performance improvements that engineers and inventors make in all technologies routinely. The results of this analysis should reduce doubt that lunar propellant mining can out-compete terrestrial-launched propellant with absolute advantage both in the near term and in the long term.

This model assumed launch costs will descend to the lowest values considered possible, and still lunar propellant gains absolute advantage. The size of the propellant market corresponding to the launch cost was modeled both pessimistically and optimistically, and in all cases the cost of lunar propellant dropped faster than launch cost and demonstrated there is a viable business case. The model made many conservative assumptions, such as neglecting aerobraking, stoichiometric fuel ratios, lunar sling-launch, and other transportation alternatives that will give lunar resources an extra advantage, and by assuming that lunar industry's firm-level economies of scale may "shut off" at a rather low throughput.

Technological maturation is needed to prove adequate reliability can be achieved for long operation in lunar conditions. This appears the be the most significant source of doubt within the aerospace community, so a sufficient demonstration with prototypes in a relevant environment may be needed. National space agencies can play a strong role in the startup of this industry by funding this effort.

Mistakes were made in some of the earlier studies. First, point-designs of a single technology were used to claim that lunar propellant cannot out-compete terrestrial propellant. Since everything depends on the technology parameter $\phi$, which can vary by orders of magnitude between different technologies and different designs, it is illogical to use one point-design of one technology to infer the fate of the entire economic sector. Second, the evolution of cost over time was found to be significant within timescales relevant to planning space activities, but prior studies focused on the performance of the first plant in its first year of operation, only. Jones et al. mentioned that their look at learning curves did not affect their pessimistic prediction, but they did not document their calculation and the opposite is clearly predicted here. Third, some prior studies failed to identify $\phi$ as the crucial parameter, so they did not work creatively to innovate a system that maximizes $\phi$ but instead put together an ad hoc system and whatever it predicted was the end of the study. Bennett et al. [2] gave an example of taking one of those ad hoc systems and reengineering it to improve $\phi$ to make it profitable.

## 11. Conclusions

Lunar-derived rocket propellant can outcompete rocket propellant launched from Earth, no matter how low launch costs go. The modeling here, which was based on normative industrial data and conservative assumptions, shows lunar propellant can be cheaper than terrestrial not only close to the Moon, but also in GTO and in LEO. This can enable a commercial business case for lunar industry to boost spacecraft, which can support development of space industry with the scaling of Earth's commercial markets. The key is to develop a lunar propellant technology that has a production mass ratio $\phi \gtrsim 35$ and to avoid expensive transportation architectures for delivering the capital to the Moon. The tent sublimation method studied by Kornuta and by Sowers is estimated to have $\phi = 400$ to $550$, an order of magnitude better than needed for absolute advantage. Strip mining methods for obtaining lunar ice are probably above the cost threshold for absolute advantage in year 1 of operation, but technological innovation and maturation, government-funded development in the early stages, and/or public-private partnership will lower costs. Furthermore, only 4 to 11 years of operation should be enough for the learning curve and increased economies of scale to lower the initial cost estimates of strip mining below the threshold of absolute advantage. It is plausible that additional technologies besides strip mining and tent sublimation may be profitable. These conclusions hold over a reasonable range of economic parameters. The remaining key challenge appears to be proof of operability for years in the lunar environment and especially the dust with a sufficiently low failure rate. The technological readiness level needs to be advanced to TRL 6 including a focus on long-run reliability and a demonstration of integrated robotic autonomy with simulated Earth-Moon communications latency to address remaining doubts.

**Funding Information**

This work was supported in part by NASA Solar System Exploration Virtual Institute cooperative agreement award NNA14AB05A, "Center for Lunar and Asteroid Surface Science."

**Acknowledgements**

I am grateful to Aeneas Weckenmann (Frankfurt, Germany), who developed the analysis method for the reusable Lunar Starship including the parameter estimates. I am grateful to Ian Lange, Division of Economics and Business, Colorado School of Mines, to Akhil Rao, Department of Economics, Middlebury College, to Pierre Lionnet, Research and Managing Director, ASC Eurospace, and to Jeff Greason, CTO and Co-Founder of Electric Sky, for reading an earlier version of the manuscript, providing helpful critiques, and identifying several issues that I had failed to discuss.

# Appendix A. Table of Data.

**Table A-1.** Study-Specific Model Parameters

| | | Tent Sublimation | | Strip Mine then Sublimation | | | Strip Mine, Beneficiate | N/A |
|---|---|---|---|---|---|---|---|---|
| | Technology: | | | | | | | |
| **Parameter** | **Units** | **K** | **S** | **CD** | **J** | **B** | **M** | **Baseline** |
| Mass of LS capital | t | 30 | 17.6 | 20.94 | 214.8 | 26.84 | 2.5 | 25 |
| Mass of space segment (RLL/OTV) | t | 5.86[1] | 5.86[1] | 5.096[2] | 22.53 | 4.5 | 1.322 | 5 |
| Payload and propellant capacity of space segment | t | 41[3] | 41[3] | 22 | 45 | 40.5 | 9.3 | 35 |
| Inert Mass Fraction | – | 0.125[3] | 0.125[3] | 0.188 | 0.26 (RLL) 0.13 (OTV) | 0.10 | 0.124 | 0.10 |
| Specific development cost, $\zeta_D$ | $K/kg | 90.7[5] | 50.2 | 149.6[4] | 122[6,8] | 191[6] | 117[6,9] | 120 |
| Specific fabrication cost, $\zeta_F$ | $K/kg | 22.7[5,6] | 41.9[7] | 60.8[4] | 30.5[6,8] | 47.6[6] | 29.4[6,9] | 40 |
| Buildup Time | y | 5[8] | 5.5 | 8 | 5[8] | 5[8] | 3[8] | 5 |
| Annual operations cost | $M | 23.9 | 78.6 | 53.4 | 158.2[10] | 20.9[10] | 10.0 | 50 |
| Operational life | y | 10 | 10 | 10 | 14 | 10 | 5 | 10 |
| Yearly mass of product at LS | t | 1640 | 1100 | 69.1 | 393.33 | 190[11] | 27.9 | 500 |

Notes:
All costs have been converted to 2022 US dollars.
(1) DTAL in Fig. 9 of Kutter et al. [121].
(2) Inferred/calculated relying on additional calculations by Shisko [8].
(3) Assumes RLL cost from B
(4) assuming 80% of hardware cost is development and 20% is fabrication
(5) Includes fabrication of RLL prorated from B. Development of RLL will be a sunk cost.
(6) Approximated due to difficulty reconstructing the figures of J.
(7) Assumes $1,684M/kg obtained from Global Security cost estimating tool [122].
(8) Not given. Estimated here.
(9) Not given. Estimated here based on much smaller system than the others.
(10) Not given. Assumes value from K prorated by capital mass.
(11) Capacity of the system, though the demand is given as 166.44 t/y.